\documentclass[twocolumn,showpacs,preprintnumbers,amsmath,amssymb]{revtex4}
\usepackage{amsmath}
\usepackage{amssymb}
\usepackage{amsfonts}
\usepackage{graphics}
\usepackage{dcolumn}% Align table columns on decimal point
\usepackage{bm}% bold math
\def\beq{\begin{equation}}
\def\eeq{\end{equation}}
\def\Schrodinger{Schr\"odinger}

\begin{document}

\title{Quantitative Relativistic Effects in the Three-Nucleon Problem}
\author{B. D. Keister}
\email{bkeister@nsf.gov}
\affiliation{%
Physics Division, \\
National Science Foundation, \\
4201 Wilson Blvd., \\
Arlington, VA 22230
}
\author{W. N. Polyzou}
\email{polyzou@uiowa.edu}
\affiliation{%
Department of Physics and Astronomy, \\
The University of Iowa, \\
Iowa City, IA 52242
}
\date{\today}% It is always \today, today,

\begin{abstract}
  The quantitative impact of the requirement of relativistic
  invariance in the three-nucleon problem is examined within the
  framework of Poincar\'e invariant quantum mechanics.  In the case of
  the bound state, and for a wide variety of model implementations and
  reasonable interactions, most of the quantitative effects come from
  kinematic factors that can easily be incorporated within a
  non-relativistic momentum-space three-body code.
\end{abstract}


\maketitle

\section{Introduction}
\label{sec:intro}

Advances in computational algorithms as well as hardware speed have
enabled a new era of precision calculations in the few-nucleon
sector.  It is now possible to compute binding properties, and in some
cases other properties, of nuclei up through A=10, starting with the
Schr\"odinger equation plus phenomenological two- and three-body
interactions, to a precision that allows for a meaningful comparison
between theory and experiment.

Computations are now precise enough that the quantitative role of
special relativity becomes relevant.  In principle, Poincar\'e
invariance is an exact symmetry that should be satisfied by all numerical
calculations, thereby permitting comparisons between theory and
experiment to rest entirely upon the nuclear dynamics.  In practice,
\begin{enumerate}
\item Consistent relativistic computation is much more numerically
  intensive, thus motivating non-relativistic calculations in practice.
\item Most estimates of relativistic effects have been quantitatively
  small enough that the non-relativistic calculations have satisfactory
  precision.
\end{enumerate}
An additional complication is that a non-relativistic model does not
imply a unique relativistic extension.  Relativistic invariance
requires the invariance of probabilities in different inertial
coordinate systems; this symmetry can be satisfied in a variety of
frameworks and models that share the same non-relativistic limit.
Since information is lost in taking the non-relativistic limit, there
is no unique way to get it back.

At the two-body level, calculations in a single inertial coordinate
system are not constrained by Poincar\'e invariance; it only ensures
that the results in all other inertial coordinate systems are
equivalent.  This affects how the two-body dynamics is embedded in the
three-body problem.  In addition, there is a unitary group of
three-body scattering equivalences that preserves Poincar\'e
invariance and leaves the two and three-body scattering and bound
state observables unchanged, at the expense making off-shell
modifications and modifications to the two and three-body
interactions.  This freedom is not constrained by experiment.  This
means that the contributions of relativistic effects, off-shell
effects, and three-body interactions cannot be uniquely separated in
the three-body problem.

Therefore, there is no single, unique ``relativistic effect'' in
few-nucleon calculations.  However, we find that, for a variety of
dynamical input and some variation in model assumptions, there is a
consistent pattern in the quantitative aspects of relativity in the
calculations.

There are two primary approaches for modeling relativistic few-body
problems in quantum mechanics.  We consider here a class of Poincar\'e
invariant quantum models~\cite{wigner} of few-particle systems.  In these
models Poincar\'e invariance is an {\it exact} symmetry that is realized by
a unitary representation of the Poincar\'e group on the few-particle
Hilbert space.  A number of equivalent representations of Poincar\'e
invariant few-body models are given in~\cite{bt,kpreview}.  Poincar\'e
invariant quantum mechanics has as its starting point a Hilbert space
with a fixed number of particles (nucleons) and a set of 2-, 3-,\dots\
body interactions.  These features it shares with its non-relativistic
counterpart based upon the Schr\"odinger equation.  It differs from
the latter in the relationship between interactions in different inertial
reference frames.  At the level of two-body phenomenology, one can
make connections to parameters fitted to data on the basis of the
Schr\"odinger equation.  If this is done in one inertial coordinate system, 
the Poincar\'e invariance can be used to generate interactions in 
any other inertial frame. 
There are several ways to do this, as will be
discussed below.  

Gl\"ockle, Kamada, and
collaborators~\cite{gloeckle1,gloeckle2,gloeckle3} have studied
aspects of Poincar\'e invariant quantum mechanics in two- and
three-body problems.  They have introduced a mapping between the
interactions in the relativistic two-body mass operator and the
interactions in the non-relativistic center-of-momentum two-body
Hamiltonian, and have solved the three-body problem for specific
interactions \cite{gloeckle4}.  Different methods for utilizing realistic
nucleon-nucleon interactions in relativistic calculations
were given by Coester, Pieper and Serduke \cite{cps} and by Gl\"ockle,
Lee, and Coester \cite{glc}.  Our goal here is to identify the
dominant sources of relativistic effects for a variety of two-body
mappings and interactions.

The second approach to the relativistic few-body problem uses
quasipotential equations.  These are relations between covariant
amplitudes in local field theory.  When some of the amplitudes are
treated as input, these relations become equations for the remaining
amplitudes.  Matrix elements of few-body observables in eigenstates of
the four momentum can be calculated from the solution of the
quasipotential equations using Mandelstam's
method~\cite{mandelstam,huang}.  For systems of strongly interacting
particles, the input to these equations is not known and must be
modeled, as in Poincar\'e invariant quantum mechanics.  While
calculations of the same observables can be computed in both
approaches, there exist no unique relation between quasipotential
equations and Poincar\'e invariant quantum mechanics.  Nevertheless,
quasipotential models have played a historically important and
valuable role in motivating the structure of model nucleon-nucleon
interactions.  The nature of the relativistic effects depends upon the
specific assumptions used to extract quantum mechanical interactions
from the quasipotential equations.  Stadler and Gross~\cite{stadler}
have studied solutions to the three-nucleon bound state problem in the
context of the spectator approximation to a meson-nucleon field
theory.  Sammarruca and Machleidt~\cite{sammarruca} have studied
specific kinematic factors that arise in extracting quantum mechanical
interactions from quasipotential equations in an effort to identify
the dominant sources of relativistic effects.

In this paper, we attempt to identify the important quantitative
relativistic effects within the framework of Poincar\'e invariant
quantum mechanics.  The goal in this paper is not to provide complete
solutions to a set of three-body problems, but rather to understand
where the major relativistic effects occur.  In fact, the most
important effects are embedded in multiplicative kinematic factors,
which makes it relatively easy to incorporate them into a
Schr\"odinger-based momentum-space three-body code.

\section{The Two-Body Problem}
\label{sec:2body}
One-nucleon momentum/spin eigenstates satisfy the normalization condition:
\begin{equation}
\label{eq:BA}
  \langle {\bf p}',
  \mu' | {\bf p} ,\mu \rangle = \delta ({\bf p}' - {\bf p})\delta_{\mu' \mu}.
\end{equation}
These states transform as mass $m$ spin $1/2$ irreducible
representations of the Poincar\'e group \cite{kpreview}.  For the
two-body problem it is useful to use a basis that transforms
irreducibly with respect to the tensor product of two one-body
representations.  These Poincar\'e irreducible eigenstates have the
structure
\begin{equation}
\label{eq:BB}
\vert {\bf P},j,\mu,k ; l ,s \rangle 
\end{equation}
where ${\bf P}$ is the total linear momentum, $j$ is the canonical spin,
$\mu$ is the $z$-component of the canonical spin, and $k$ is related to 
the invariant  
invariant mass of the non-interacting two-body system by
\begin{equation}
M^{(0)} = 2 \sqrt{m^2 + k^2}. 
\label{eq:BC}
\end{equation}
The quantum numbers $l$ and $s$ are degeneracy quantum numbers that 
determine the multiplicity of each representation with the same
$M^{(0)}$ and $j$; $l$ and $s$ have the same
spectrum as the non-relativistic total spin and orbital angular 
momentum quantum numbers.  The overlap coefficients with the 
tensor product of single 
particle states are Poincar\'e Clebsch-Gordan coefficients which can be 
found in (\cite{joos}\cite{coester}\cite{kpreview}\cite{stora}).

The two-body mass operator is defined by adding interactions to the 
non-interacting mass operator:
\begin{equation} 
M= M^{(0)}+v .
\label{eq:BD}
\end{equation}
In Poincar\'e invariant quantum mechanics the delta functions that 
multiply interactions are important.  The 
two-body interaction acting on the two-body 
Hilbert space is denoted by $v$;  $\bar{v}$ denotes the internal two-body 
interaction related to $v$ by 
\[
\langle {\bf P}', j' , \mu', k', l' ,s' \vert v 
\vert {\bf P}, j , \mu, k, l ,s \rangle =
\]
\begin{equation}
\delta ({\bf P}'-{\bf P}) \delta_{j'j}\delta_{\mu' \mu}
\langle k',l',s' \vert \bar{v}^j \vert k,l,s \rangle .
\label{eq:BE}
\end{equation} 

With this choice of $v$ it is possible to find simultaneous
eigenstates of $M$, ${\bf P}$, $j^{(0)}$, and ${\bf j}^{(0)} \cdot
\hat{\bf z}$.  These states transform like mass $M$, spin ${\bf
j}^{(0)}$ irreducible representations of the Poincar\'e group.  Since
these eigenstates are complete they define the interacting two-body
dynamics.

Two-particle scattering is described by the transition 
operator
\begin{equation}
{t}(e) = {v}+ {v} {1\over {e +i0 -M  }} {v}
\label{eq:BF}
\end{equation}
which satisfies the relativistic Lippmann-Schwinger
equation:
\begin{equation}
{t}(e) = {v} + {v} 
{1\over {e+i0 - M^{(0)} }} {t}(e).
\label{eq:BG} 
\end{equation}
This transition operator has the same relation to the differential
cross section as the corresponding non-relativistic expression, except
that the single particle masses, $m$, appearing in the incident
current and phase space factors are replaced by the corresponding
single particle energies, $\omega=\sqrt{k^2 + m^2}$.  A proof based on
time-dependent scattering theory is given in \cite{kpreview}.

The reduced two-body transition operator $\bar{t}(e)$ is related to $t(e)$ by
\[
\langle {\bf P}', j' ,\mu',k' , l' s'  | t(e) 
| {\bf P}, j , \mu, k, l ,s \rangle
= 
\]
\begin{equation}
\delta ({\bf P}' - {\bf P})\delta_{j'j} \delta_{\mu' \mu}
\langle k',l',s' \vert \bar{t}^j(e) \vert k,l,s \rangle .
\label{eq:eq:BH}
\end{equation}

\section{Connections to Two-Body Phenomenology}
\label{sec:phenomenology}

All realistic descriptions of the quantum mechanical two-nucleon
system utilize adjustable parameters in the interactions to fit
published nucleon-nucleon phase-shift data and the deuteron binding
energy.  For Poincar\'e invariant quantum mechanical models, these
parameters involve interaction strengths and ranges; for
quasipotential models, the parameters may include meson-nucleon
coupling strengths and form factors.  Since there is an extensive
investment of effort in fitting nucleon-nucleon data to solutions to
the Schr\"odinger equation, it is advantageous to directly use existing
high-quality interactions to construct equivalent relativistic
interactions.

The relationship between relativistic and non-relativistic models fit
to the same data is more complicated than the relation obtained by
taking the non-relativistic limit of the relativistic model.
Understanding the nature of the fitting process is the first element
needed to understand the nature of relativistic corrections.  The
scattering cross section is a relativistic invariant~\cite{Moller};
however, the angular distributions and spins have a frame dependence
with known kinematic transformation properties.  Kinematic Lorentz
transformations are used to correctly transform measured laboratory
differential scattering cross sections to the center of momentum
frame.  The properly transformed data are used to fit interactions
that reproduce this data by solving the non-relativistic Schr\"odinger
equation.

The scattering operator and consequently the phase shifts for a given
angular momentum are Poincar\'e invariant.  The phase shifts can be
tabulated as functions of the invariant center of momentum momentum or
energy, $\delta_j (k)$ or $\delta'_j (e)$. 
These functions are equal when the center of momentum energy $e$ and 
invariant momentum $k$ are related by the correct relativistic 
relation.  For equal mass particles this relation is
$e=2\sqrt{k^2+m^2}$ which implies 
\begin{equation}
\delta_j (k) = \delta'_j (2\sqrt{m^2 + k^2}) 
\qquad
\delta'_j (e) = \delta_j \left (\sqrt{{e^2 \over 4} -m^2} \right ).
\label{eq:CA}
\end{equation}

In the fitting procedure 
these phase shifts are identified with phase shifts obtained by
solving the Schr\"odinger equation.  This can be done in one of two {\it
inequivalent} ways \cite{allen}.  The first is by identifying $\delta_j (k)$ with
the corresponding $k$-dependent Schr\"odinger phase shift and the
second is by identifying $\delta_j' (e)$ with the corresponding
$e$-dependent Schr\"odinger phase shift.  These are inequivalent
because non-relativistically $e$ and $k$ are related by 
$e=k^2/m + 2m$.

If these procedures are used there are no relativistic corrections in
the center of momentum frame when compared to a relativistic model fit
to the same data.  The Schr\"odinger equation produces the exact
Poincar\'e invariant phase shifts either as functions of the center of
momentum energy or momentum (but not both). 

The problem that can arise is if the phase shifts are fit to
transformed cross section data as functions of $k$ and the potential
is constructed by matching the phase shifts obtained from the
Schr\"odinger equation as functions of $e$, then the phase shift or
cross section predicted at energy $e=k^2/m + 2m$ will be the same as
the ``measured'' phase shift a shifted energy $e'=2 \sqrt{m^2 + k^2}$.
Similarly, if the phase shifts are fit to transformed cross section
data as functions of $e$ and the potential is constructed by matching
the phase shifts obtained from the Schr\"odinger equation as functions
of $k$ then the phase shift or cross section predicted at momentum $k=
\sqrt{em-2m^2}$ will be the same as the measured phase shift a shifted
momentum $k'=\sqrt{e^2/ 4-m^2}$.

This inconsistency increases with momentum.  At relativistic energies
the inconsistencies are of the same scale as typical relativistic
corrections.  This is illustrated in Figure 1, which compares the
non-relativistic and relativistic energies $e_{nr}(k)$ to $e_r(k)$ for
a nucleon with momenta $k$ up to 1~GeV.  These curves show the scale
of the energy mismatch as a function of $k$; it is a 7\% effect at
1~GeV.  Figure 2 illustrates the same effect with the total cross
section.  The curves in Figure 2 compare the total cross
sections in the Born approximation for the Malfliet-Tjon V potential
\cite{malfliet}.  The curve in Figure 2 is the ratio $\sigma (k) / \sigma
(k\sqrt{1+k^2/4m^2})$ for $k$ up to 1~GeV/$c$.  In this example
figure shows that an inconsistent treatment of the phase equivalence,
while small at low momenta, leads to a 15\% percent error in the total
cross section at 1~GeV/$c$.

\begin{figure}
\begin{center}
\rotatebox{270}{\resizebox{2.9in}{!}{
\includegraphics{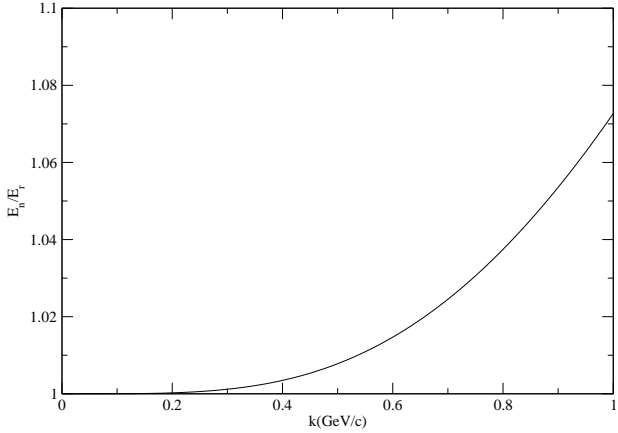}}
}
\end{center}
\caption{Relativistic vs Non-relativistic CM energy as a function of
$k$.
}
\end{figure}

%Figure 2 shows the ratio of the total cross section in the Born approximation
%when the cross section is identified with the non-relativistic cross section
%is identified with the physical cross section as a function of momentum 
%to the cross section obtained by making the identification as a function 
%of energy  as a function of momentum of a function of energy .  Both cross 
%sections are plotted as functions of momentum.

\begin{figure}
\begin{center}
\rotatebox{270}{\resizebox{2.9in}{!}{
\includegraphics{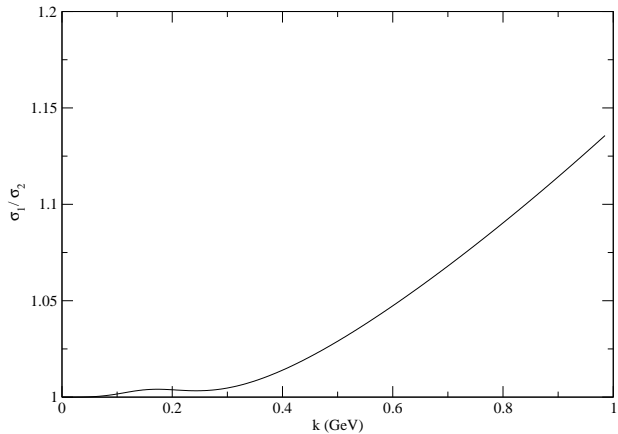}}
}
\end{center}
\caption{Differences in invariant cross section identified as
functions of relative momentum to relative energy.  
}
\end{figure}

Thus, the first question that needs to be considered is how the
non-relativistic interaction was constructed.  For Argonne $V18$,
which is a typical example of a realistic nucleon-nucleon interaction,
the phase shifts are determined as functions of laboratory beam 
energy \cite{wiringa} which is converted to center-of-momentum momentum 
using 
\begin{equation}
- p_b \cdot p_t = e_{Lab} m = (k^2 + m^2 ) + k^2 .
\label{eq:CC}
\end{equation} 
Thus, for the $AV18$ interaction, the solution of the Schr\"odinger 
equation reproduces the invariant phase shifts $\delta_j (k)$ as
functions of the invariant momentum.

The methods discussed in this paper make use of the phase 
equivalence between the interactions in the relativistic and
non-relativistic Schr\"odinger equations.  We consider 
three ways of utilizing two-body phenomenology based upon the
fitted interactions in \Schrodinger\ equation in Poincar\'e 
invariant quantum mechanics.  These methods relate the half-shell
relativistic two-body transition matrix elements to the 
half-shell non-relativistic transition matrix elements under
different sets of assumptions about the phase shift fitting.

\subsection{Coester-Pieper-Serduke (CPS)}

This method was formulated to study relativistic effects in nuclear
matter.  It uses a nucleon-nucleon interaction, such as $AV18$, fit to
$\delta_j (k)$ to construct a scattering equivalent relativistic
interaction.

CPS~\cite{cps} add an interaction to the square of the non-interacting 
mass operator to get a \Schrodinger-like equation: 
\begin{equation}
\label{eq:DA}
M^2 = M^{(0)2} + {u} = 4m h
\qquad h:= {k^2\over m} + {{u}\over 4m}+m  .
\end{equation}
The operator ${u}/4m = {v}_{nr}$ is identified with the 
interaction that appears in the non-relativistic Schr\"odinger Hamiltonian,
$h$.

The relationship between ${u}$ in Eq.~(\ref{eq:DA}) and 
the relativistic interaction ${v}$ in
Eq.~(\ref{eq:BE}) is
\begin{equation}
  \label{eq:DB}
  {u} = {v}^2 + \{M^{(0)}, {v}\}.
\end{equation}
It would be tedious to evaluate ${v}$ directly;  however,
it is never needed because $M$, $M^2$ and $h=k^2/m + {v}_{nr}$, all have the 
same eigenstates. 

If ${\hat t}(e)$ is generated by the ``Lippmann-Schwinger''
equation
\begin{equation} 
{\hat t} (e) = {u} + {u} {1 \over e^2 - M^{(0)2} + i0^+} 
\hat{t}(e),
\label{eq:DC}
\end{equation}
and $\vert \psi_k^+ \rangle$ is a scattering eigenstate with momentum
$k$ and $\vert \psi_k \rangle$ is the corresponding plane wave state,
then the following half-shell relations follow:
\[
{t}(e(k)) \vert \psi_k  \rangle  = 
{v} \vert \psi_k^+  \rangle =
(M-M_0 ) \vert \psi_k^+  \rangle =
\]
\[ 
{1 \over e(k) + M_0} (M^2 -M_0^2 )
\vert \psi_k^+  \rangle =
\]
\[
{1 \over e(k) + M_0} {u} \vert \psi_k^+  \rangle =
{1 \over e(k) + M_0} \hat{t}(e(k)) \vert \psi_k  \rangle =
\]
\begin{equation}
{4m \over e(k) + M_0} {v}_{nr} \vert \psi_k^+  \rangle =
{4m \over e(k) + M_0} {t}_{nr}(e_{nr}(k)) \vert \psi_k  \rangle .
\label{eq:DD}
\end{equation} 
Taking matrix elements in the irreducible plane wave states
$\vert k,j,l,s \rangle$ gives the relation
\[
\langle k',l',s' \vert \bar{t}^j (e(k)) \vert k,l,s \rangle =
\]
\[
[e(k) + e(k'))]^{-1}
\langle k',l',s' \vert {\hat t}^j  (e(k)) \vert k,l,s \rangle =
\]
\begin{equation}
{4m \over 
e(k) + e(k')}
\langle k',l',s' \vert {\bar t}_{nr}^j  (k^2/m) \vert k,l,s \rangle 
\label{eq:DE}
\end{equation}
where $e(k)=2 \omega (k) = 2\sqrt{k^2 + m^2}$,  $e_{nr}(k)=k^2/m$, and 
$\bar{t}^j$ is the reduced non-relativistic two-body transition 
operator. 

This gives the desired relation between the relativistic and
non-relativistic half off-shell transition matrix elements.  This
formula will be used to construct the kernel of the three-body
equations.
  
It has been shown that for the CPS method the relativistic and
non-relativistic cross sections are identical functions of the 
invariant momentum $k$ \cite{kpreview}, 
and the relativistic and non-relativistic 
bound states have the same bound-state wave numbers. This phase 
equivalence is exact.
 
\subsection{Gl\"ockle-Lee-Coester (GLC)}

Gl\"ockle,Lee and Coester introduced an approximate variation of the
CPS method that did not require introducing the interaction ${u}$.
This approximation was used in the first calculation of a
three-body binding energy based on Poincar\'e invariant quantum
mechanics \cite{glc}.

Like the CPS method, the GLC method uses a nucleon-nucleon
interaction, such as $AV18$ \cite{wiringa}, fit to $\delta_j (k)$ to
construct a scattering equivalent relativistic interaction.  In this
case the scattering equivalence is only approximate.
 
GLC~\cite{glc} define a new interaction:
\begin{equation}
  \label{eq:EA}
\langle k',l',s' \vert {\bar v}_{nr}^j \vert  k,l,s \rangle
  = \sqrt{\omega'\over m} 
\langle k',l',s' \vert \bar{v}^j \vert k,l,s \rangle 
\sqrt{\omega\over m},
\end{equation}
with the corresponding $t$ matrix elements:
\[
\langle k',l',s' \vert {\bar t}^j_{nr} (k^2/m) \vert  k,l,s \rangle
:=
\]
\begin{equation}
\sqrt{\omega'\over m} 
\langle k',l',s' \vert \bar{t}^j (e(k)) \vert k,l,s \rangle
\sqrt{\omega\over m}.
\label{eq:EB}
\end{equation}

Equation (\ref{eq:EB}) replaces (\ref{eq:DE}) in the CPS 
method.  With the above definitions if $\bar{t}$ satisfies
the relativistic Lippmann-Schwinger equation, it 
does not follow that  $\bar{t}_{nr}$ will satisfy the 
non-relativistic Lippmann
Schwinger equation, however is straightforward to derive   
\[
\langle k',l',s' \vert {\bar t}_{nr}^j (k^2/m) \vert  k,l,s \rangle = 
\langle k',l',s' \vert{\bar v}_{nr}^j \vert k,l,s \rangle
  + 
\]
\[
\sum_{l''s''} \int k^{\prime\prime 2} dk'' \,
{\langle k',l',s' \vert {\bar v}^j_{nr} \vert k'',l'',s'' \rangle
\over
k^2/m - k''{}^2/m +i0} \times
\]
\begin{equation}
\langle k'',l'',s'' \vert{\bar t}^j_{nr}(k^2/m) \vert k,l,s \rangle
   \left[1 + {\omega - \omega'' \over 2\omega''}\right].
\label{eq:EC}
\end{equation}
The CPS approximation is equivalent to 
neglecting the term, ${\omega - \omega'' \over 2\omega''}$,
which is zero on shell.  Neglecting this term
gives the non-relativistic Lippmann-Schwinger 
equation.  CPS approximate the solution, $\bar{t}^j_{nr}$,
of (\ref{eq:EC}) by the 
non-relativistic transition operator.

It can be shown that this method does not lead to an exact phase
equivalence, it is possible to construct systematic corrections that
converge to the $CPS$ result.  In the original CPS calculation this
approximation was found to be accurate. It was improved using a
slight adjustment of the interaction parameters.

\subsection{Gl\"ockle-Kamada (GK)}

The GK approach is designed to produce a relativistic dynamical model
that is phase equivalent to a corresponding non-relativistic 
model where the phases shifts are fit as functions of center 
of momentum energy, $\delta' (E)$.  

The GK~\cite{gloeckle1} approach uses a unitary rescaling of the momentum
variables to change the non-relativistic kinetic energy into the 
relativistic kinetic energy.

For each value of the momentum $k$, GK define the momentum $q$ 
by identifying the relativistic and non-relativistic energies
\begin{equation}
  \label{eq:FA}
  2m + {q^2\over m} = 2 \sqrt{m^2 + k^2}.
\end{equation}
This can solved for be either for $k(q)$ or $q(k)$.
%\begin{equation}
%  \label{eq:FB}
%  k = q\sqrt{1 + {q^2\over 4 m^2}};
%\end{equation}
%\begin{equation}
%  \label{eq:FC}
%  q = \sqrt{2m}\sqrt{\omega_k - m}.
%\end{equation}
Defining $h(q)$ as
\begin{equation}
 \label{eq:FB}
  h^2(q) := {q^2 \over k^2} {dq \over dk} 
\end{equation}
Gl\"ockle and Kamada identify a ``relativistic'' interaction:
\[
\langle k',l',s' \vert {\bar v}^j \vert  k,l,s \rangle
:=
\]
\begin{equation} 
h(k') \langle q(k'),l',s' \vert \bar{v}^j_{nr} 
\vert q(k),l,s \rangle  h(k)
\label{eq:FC}
\end{equation}
and $t$ matrix:
\[  
\langle k',l',s' \vert {\bar t}^j (e(k)) \vert k,l,s \rangle
:=
\]
\begin{equation}
h(k') \langle q(k'),l',s' \vert \bar{t}_{nr}^j (e_{nr} (q(k)) \vert
q(k),l,s \rangle h(k).
\label{eq:FD}
\end{equation}
Then $\bar{t}^j$ and ${\bar t}^j_{nr}$ satisfy the relativistic 
and non-relativistic Lippmann-Schwinger equations, 
as functions of $k$ or $q$, respectively. 

With this method (\ref{eq:FD}) replaces 
(\ref{eq:EB}) or (\ref{eq:DE}).

In \cite{gloeckle1}
Kamada and Gl\"ockle show that with the KG method the relativistic and
non-relativistic phase shifts are identified as functions of the 
invariant energy $E$, and the relativistic and non-relativistic 
bound states have identical binding energy.

\section {The Three-Body Problem}
\label{sec:3body}

To construct kinematic variables for the three-body system, let
$M^{(0)}$ denote the invariant mass of the non-interacting three-body
system and $M^{(0)}_{ij}$ be the invariant mass for the
non-interacting two-body sub-system consisting of particles $i$ and $j$.

Plane-wave basis states are tensor products of one-body Poincar\'e 
irreducible representation space basis vectors 
\begin{equation}
\vert {\bf p}_1, \nu_1, {\bf p}_2, \nu_2, {\bf p}_3, \nu_3 \rangle . 
\end{equation}

As in the non-relativistic case it is useful to 
define the total momentum
\begin{equation}
{\bf P}:= \sum_{i=1}^3 {\bf p}_i 
\label{eq:GA} 
\end{equation}
and relativistic Jacobi momenta, 
which are the the three-vector components of 
\begin{equation}
K_{ij}^{\mu} := L^{-1}({{\bf P} \over M^{(0)}})^{\mu}{}_{\nu}  p_k^{\nu}  
\label{eq:GB}
\end{equation}
\begin{equation}
k_{ij}^{\mu} := L^{-1}({{\bf p}_i + 
{\bf p}_j \over M^{(0)}_{ij} })^{\mu}{}_{\nu} 
p_i^{\nu} 
\label{eq:GC}
\end{equation}
where $L(\cdot )$ denotes a rotationless (canonical) Lorentz boost.
These choices are appropriate for an ``instant-form'' representation.
We use this representation for the purpose of illustration.  Other
choices give similar formulas.

The non-relativistic Jacobi moment are obtained if the Lorentz
boosts $L$ are replaced by the corresponding Galilean boosts.  Both 
the relativistic and corresponding Jacobi momentum operators have 
identical spectra. 

For the three-body problem it is useful to successively pairwise 
couple the one-body Poincar\'e irreducible representations to 
obtain three-body Poincar\'e irreducible basis vectors.
This is done using pairs of Poincar\'e Clebsch-Gordan coefficients.
There are three bases that differ in the order of the pairwise 
coupling.  We use the following shorthand notation for the basis 
that couples the irreducible representation associated with the 
pair $(kl)$ to the representation associated with the ``spectator''
particle, $m$:
\begin{equation}
\vert {\bf P}, j , \mu, (kl)(m) \rangle :=
\vert {\bf P}, j , \mu ; L_{kl}, S_{kl}, K_{kl}, j_{kl}, l_{kl}, s_{kl}, k_{kl}
\rangle . 
\label{eq:GD}
\end{equation}
Here $K_{kl}$ and $k_{kl}$ are magnitudes of the relativistic Jacobi 
momenta.  They are related to the invariant masses of the two- and 
three-body irreducible 
representation by 
\begin{equation} 
M^{(0)}_{kl} = 2 \sqrt{{ k}_{kl}^2 + m^2}
\label{eq:GE}
\end{equation}
and 
\begin{equation}
M^{(0)} 
= \sqrt{M^{(0)}_{kl}{}^2 + { K}_{kl}^2} + \sqrt{m^2+{K}_{kl}^2}.
\end{equation} 

The two-body interaction $V_{kl}$ and reduced two-body interaction 
$\bar{V}_{kl}$ are defined in this basis by
\[
\langle  {\bf P}, j , \mu, (kl)(m) \vert V_{kl} \vert
{\bf P}', j' , \mu', (kl)(m)' \rangle =
\]
\begin{equation}
\delta ({\bf P} - {\bf P}\,')\delta_{jj'} \delta_{\mu \mu'} 
\langle (kl)(m) \vert {\bar V}_{kl} \vert
(kl)(m)' \rangle 
\label{eq:GF}
\end{equation}
where
\[
\langle (kl)(m) \vert {\bar V}_{kl} \vert
(kl)(m)' \rangle =
{\delta (K_{kl} - K_{kl}') \over K_{kl} K_{kl}'} \times
\]
\begin{equation}
\delta_{L_{kl}L_{kl}'} \delta_{S_{kl}S_{kl}'}
\delta_{j_{kl}j_{kl}'}
\langle k_{kl}, l_{kl}, s_{kl} \vert \bar{v}^{j_{kl}} \vert
k_{kl}', l_{kl}', s_{kl}' \rangle .
\label{eq:GG}
\end{equation}
and $\langle \cdots  \vert \bar{v}^{j_{kl}} \vert
\cdots  \rangle$ is the reduced relativistic two-body interaction.
The two-body transition operators can be embedded in the three-particle 
Hilbert space:
\[
\langle (kl)(m) \vert {\bar T}_{kl}(e_{kl}) \vert
(kl)(k)' \rangle :=
{\delta (K_{kl} - K_{kl}') \over K_{kl} K_{kl}'} \times
\]
\begin{equation}
\delta_{L_{kl}L_{kl}'} \delta_{S_{kl}S_{kl}'}
\delta_{j_{kl}j_{kl}'}
\langle k_{kl}, l_{kl}, s_{kl} \vert \bar{t}^{j_{jk}} (e_{kl}) \vert
k_{kl}', l_{kl}', s_{kl}' \rangle .
\label{eq:GH}
\end{equation}

It is important to note that the interaction $\bar{V}_{kl}$ is
constructed from $\bar{v}_{kl}$ rather that $v_{kl}$. This is
intentional; $\bar{V}_{kl}$ commutes with ${\bf K}_{kl}$ rather than
${\bf p}_k + {\bf p}_l$.  It cannot commute with both because the
relation (\ref{eq:GB}) between these variables involves $k_{kl}$,
which does not commute with $v_{kl}$.  This requirement is used to
establish three-body Poincar\'e invariance.  It violates cluster
properties for the generators, but cluster properties are still valid
for the scattering matrix \cite{coester}.

The two-body interactions that appear in the Faddeev-Lovelace equations
have a complex relation to the interaction $\bar{V}_{kl}$ introduced above.
%This relation is determined embedding two body mass 
%operator is embedded in the three particle Hilbert space 
%by adding them to the free two-body mass operator and then using 
%that in the three body mass operator.  
%
%where $M^{(0)}_{12}$ is the non-interacting mass of the $(12)$
%subsystem,
%\begin{equation}
%M^{(0)}_{12} = 2\sqrt{m^2 + {\bf k}_{12}^2} 
%\end{equation}
%and
%\begin{equation}
%K_{12}^\mu 
%= L^{-1} ({{\bf P}\over M^{(0)}})^\mu{}_\nu
%P_{12}^\nu \label{COA}
%\end{equation}
%is operator whose eigenvalue is the three-momentum of the $(12)$ 
%cluster in the three-body rest frame, evaluated using a kinematic
%rotationless Lorentz transformation.  

%The total spin is constructed by combining the spins of the two-body
%subsystem with the third spectator:
%\begin{equation}
%{\bf j} = {\bf j}_{12} + {\bf j}_{3}
% + {\bf L},\label{COD}
%\end{equation}
%where ${\bf L}$ is the internal orbital angular momentum,
%and ${\bf k}_3 = -{\bf K}_{12}$.

The first step in the derivation of the interaction terms in 
the Faddeev-Lovelace equations 
is to add the interaction ${V}_{jk}$ to the $(jk)$ subsystem in
the presence of a third particle spectator, $l$.  The mass operator 
$M_{(jk)(l)}$ is defined 
by its matrix elements in the basis (\ref{eq:GD}) as
\[
\langle  {\bf P}',j',\mu',(jk)(l)' | 
M_{(jk)(l)}| {\bf P}, j, \mu, (jk)(l) \rangle =
\]
\begin{equation}
\delta({\bf P}' - {\bf P})\delta_{j'j} \delta_{\mu' \mu} 
\langle (jk)(l)' | 
\bar{M}_{(jk)(l)}|(jk)(l) \rangle 
\label{eq:GI}
\end{equation} 
where 
\[
\langle (jk)(l)' | 
\bar{M}_{(jk)(l)}|(jk)(l) \rangle =
\]
\[
{\delta(K_{jk}' - K_{jk})
\over K'_{jk}K_{jk}}
\delta_{L_{jk}'L_{jk}}
\delta_{S_{jk}'S_{jk}}\delta_{j_{jk}'j_{jk}} 
\langle l_{jk}' ,s_{jk}',k_{jk}' | 
\] 
\begin{equation}
\left [ \sqrt{(M_{jk}^{(0)}+\bar{V}^{j_{jk}}_{jk})^2 + K_{jk}^2} 
+ \sqrt{m^2+ K_{jk}^2}  \right ] |
l_{jk} ,s_{jk},k_{jk} \rangle.
\label{eq:GJ}
\end{equation}
where $\bar{V}^{j_{jk}}_{jk}$ is the two-body interaction introduced 
in equation (\ref{eq:GF}).

It follows from the definition (\ref{eq:GI}) that
\begin{equation}
[ M_{(jk)(l)} , {\bf j}^{(0)}]_-= 
[ M_{(jk)(l)} , {\bf P}]_-=0 ,
\label{eq:GK}
\end{equation} 
and the matrix element of $M_{(jk)(l)}$ has no explicit ${\bf P}$ dependence
in the basis (\ref{eq:GD}).

Two-body interactions $\tilde{V}_{(jk)(l)}$ that appear in the 
Faddeev-Lovelace 
equations  
are defined by
\begin{equation}
\tilde{V}_{(jk)(l)} := M_{(jk)(l)} - M^{(0)} .
\label{eq:GL}
\end{equation}
These can be used to define a three-body mass operator
\begin{equation}
M:= M^{(0)} + \tilde{V}_{(12)(3)}+ \tilde{V}_{(23)(1)} +\tilde{V}_{(31)(2)}
=  M^{(0)} + \tilde{V}.
\label{eq:GM}
\end{equation} 

It follows from the definition (\ref{eq:GI}) that
\begin{equation}
[ M , {\bf j}^{(0)} ]_-= 
[ M , {\bf P} ]_- =0,
\label{GN}
\end{equation} 
and that simultaneous eigenstates of $M$, ${\bf P}$, ${\bf j}^{(0)}$
and ${\bf j}^{(0)} \cdot \hat{\bf z}$ transform as irreducible mass
$M$ spin ${\bf j}^{(0)}$ eigenstates of the Poincar\'e group.  Since
these eigenstates are complete on the three-body Hilbert space, a
Poincar\'e invariant dynamics is defined by the requirement that these
eigenstates transform irreducibly with respect to the Poincar\'e
group.

To construct dynamical equations for the three-body transition
operators, first define residual interactions $\tilde{V}^a$ for each
two-cluster partition $a=(ij)(k)$ by
\begin{equation} 
\tilde{V}^{a} = \tilde{V} - \tilde{V}_a . 
\label{eq:GO}
\end{equation}
The three-body channel transition operators 
\begin{equation} 
T^{ab}(W) := \tilde{V}^b + \tilde{V}^a {1 \over W - M + i0^+} \tilde{V}^b
\label{eq:GP}
\end{equation}
where $W$ is a constant, 
satisfy the Faddeev-Lovelace equations
\begin{equation} 
T^{ab}(W) = \tilde{V}^b +
\sum_{c \not= a} \tilde{V}_c {1 \over W - M_c + i0^+ } T^{cb}(W).
\label{eq:GQ}
\end{equation}
where the sum runs over two-cluster partitions, $c=(jk)(l), (kl)(j)$ and 
$(lj)(k)$.
The identity 
\begin{equation}
\tilde{V}_c {1 \over W - M_c + i0^+ } =
\tilde{T}_c (W) {1 \over W - M_0 + i0^+ }
\label{eq:GR}
\end{equation}
where
\begin{equation}
\tilde{T}_c (W) := \tilde{V}_{c} + 
\tilde{V}_{c} {1 \over W-M_c + i0^+} \tilde{V}_{c}
\label{eq:GS}
\end{equation} 
leads to the equivalent form of the Faddeev-Lovelace 
equations:
\begin{equation} 
T^{ab}(W) = \tilde{V}^b +
\sum_{c \not= a} \tilde{T}_c (W) {1 \over W - M_0 + i0^+ } T^{cb}(W).
\label{eq:GT}
\end{equation}

The difficulty with the relativistic Faddeev-Lovelace equations is
that the transition operator $\tilde{T}_c (W)$ differs from the off-shell
two-body transition operators.  This can be remedied by the observing
that the eigenfunctions of $M_{12}$ and $M_{(12)(3)}$ differ by 
delta functions.  Our goal is to replace these by expressions involving 
non-relativistic transition operators. 

It follows that
on the right half shell,
\[
\langle  (jk)(l)' | \tilde{T}^j_{(jk)(l)} (W) 
| (jk)(l) \rangle =
\]
\begin{equation}
\langle  (jk)(l)' | (M_{(jk)(l)} -M_0 )
| (jk)(l)^{\pm} \rangle
\label{eq:GU}
\end{equation}
where the $)^{\pm}$ indicates a scattering eigenstate with 
invariant mass eigenvalue   
\begin{equation}
W= \sqrt{{ K}^2_{jk} + 4k^2_{jk} + 4 m^2} + 
\sqrt{{ K}^2_{jk} + m^2}
\label{eq:GV}
\end{equation}
and
\[
\langle  (jk)(l)' | \bar{T}^{j_{jk}}_{jk} (e) 
|  (jk)(l) \rangle =
\]
\begin{equation}
\langle  (jk)(l)' | (M_{jk} -M_{jk0} )
| (jk)(l)^{\pm} \rangle
\label{eq:GW}
\end{equation}
for $e= 2 \sqrt{k_{jk}^2 + m^2 }$.
Since the initial states are eigenstates of $M_{(jk)(l)}$ and $M_{jk}$ 
respectively and the final states are eigenstates of 
$M^{(0)}$ and $M^{(0)}_{jk}$ respectively, it follows that
\[
\langle  (jk)(l)' | \tilde{T}^j_{(jk)(l)} (W) 
|  (jk)(l) \rangle =
\]
\[
{2\sqrt{k_{jk}^2 + m^2} + 2\sqrt{k_{jk}^{\prime 2} + m^2}      
\over
\sqrt{K_{jk}^2 + 4k_{jk}^2 + 4m^2} +
\sqrt{K_{jk}^{\prime 2} + 4k_{jk}^{\prime 2} + 4m^2}} \times
\]
\begin{equation} 
\langle   (jk)(l)' | \bar{T}^{j_{jk}}_{jk} (e) 
|  (jk)(l) \rangle 
\label{eq:GX}
\end{equation}
for the right half shell transition matrix elements.  There is  
a similar expression for the left-half-shell transition amplitudes.

The right hand side of this equation can be expressed in term of the 
reduced half-shell two-body transition operators which are 
related to the corresponding non-relativistic transition 
operators by (\ref{eq:DE}) (\ref{eq:EB}) and (\ref{eq:FD}).
Combining these equations with (\ref{eq:GH}) gives
\[
\langle  (jk)(l)' | \tilde{T}^j_{(jk)(l)} (W) 
| (jk)(l) \rangle =
\]
\[
\delta_{L_{kl}L_{kl}'} \delta_{S_{kl}S_{kl}'}
\delta_{j_{kl}j_{kl}'}
{\delta (K_{kl} - K_{kl}') \over K_{kl} K_{kl}'}
\times
\]
\[
{2\sqrt{k_{jk}^2 + m^2} + 2\sqrt{k_{jk}^{\prime 2} + m^2}      
\over
\sqrt{K_{jk}^2 + 4k_{jk}^2 + 4m^2} +
\sqrt{K_{jk}^{\prime 2} + 4k_{jk}^{\prime 2} + 4m^2}}
\times
\]
\begin{equation}
{4m \over 
e(k_{kl}) + e(k'_{kl})}
\langle k_{kl}',l_{kl}',s_{kl}' 
\vert {\bar t}_{nr}^j  (k_{kl}^2/m) \vert k_{kl},l_{kl},s_{kl} \rangle 
\label{eq:GXA}
\end{equation}
for the CPS method,
\[
\langle  (jk)(l)' | \tilde{T}^j_{(jk)(l)} (W) 
| (jk)(l) \rangle \approx
\]
\[
\delta_{L_{kl}L_{kl}'} \delta_{S_{kl}S_{kl}'}
\delta_{j_{kl}j_{kl}'}
{\delta (K_{kl} - K_{kl}') \over K_{kl} K_{kl}'} \times
\]
\[
{2\sqrt{k_{jk}^2 + m^2} + 2\sqrt{k_{jk}^{\prime 2} + m^2}      
\over
\sqrt{K_{jk}^2 + 4k_{jk}^2 + 4m^2} +
\sqrt{K_{jk}^{\prime 2} + 4k_{jk}^{\prime 2} + 4m^2}}
\times
\]
\begin{equation}
\sqrt{\omega'_{kl}\over m} 
\langle k_{kl}',l_{kl}',s_{kl}' \vert \bar{t}^j (e(k_{kl})) \vert k_{kl},
l_{kl},s_{kl} \rangle
\sqrt{\omega_{kl}\over m}.
\label{eq:GXB}
\end{equation}
for the GLC method, and 
\[
\langle  (jk)(l)' | \tilde{T}^j_{(jk)(l)} (W) 
| (jk)(l) \rangle =
\]
\[
\delta_{L_{kl}L_{kl}'} \delta_{S_{kl}S_{kl}'}
\delta_{j_{kl}j_{kl}'}
{\delta (K_{kl} - K_{kl}') \over K_{kl} K_{kl}'} \times
\]
\[
{2\sqrt{k_{jk}^2 + m^2} + 2\sqrt{k_{jk}^{\prime 2} + m^2}      
\over
\sqrt{K_{jk}^2 + 4k_{jk}^2 + 4m^2} +
\sqrt{K_{jk}^{\prime 2} + 4k_{jk}^{\prime 2} + 4m^2}}
\times
\]
\begin{equation}
h(k_{kl}') \langle q(k_{kl}'),l_{kl}',s_{kl}' 
\vert \bar{t}_{nr}^j (e_{nr} (q(k_{kl})) \vert
q(k_{kl}),l_{kl},s_{kl} \rangle h(k_{kl})
\label{eq:GXC}
\end{equation}
for the GK method. 
Unfortunately these relations only hold for the half-shell transition 
matrix elements.  The relations (\ref{eq:GXA})-(\ref{eq:GXB}) are 
exact. 

The fully off-shell matrix elements of $\tilde{T}_c (W)$ that are
needed as input in the Faddeev Lovelace equation can be obtained from
the two-body bound state wave functions and the half-shell transition
matrix elements by quadrature, or directly from the half-shell
transition matrix elements by solving the first resolvent equation.
Either of these methods provides a means for constructing the kernel
of the relativistic Faddeev-Lovelace equations from solution of the
relativistic Lippmann-Schwinger equations.  This step, {\it of
constructing the fully off-shell input to the Faddeev Lovelace
equation}, is the most non-trivial difference between the
non-relativistic and relativistic three-body problems. It involves
either an additional spectral expansion or solving an additional
integral equation.

While the purpose of this paper is to discuss 
the consequences of avoiding this additional step, we discuss
this step briefly for completeness.

The first method for constructing the off-shell transition 
operators from the half-shell operators uses the spectral decomposition
which can be expressed compactly by
\[
\langle f' \vert \tilde{T}_{a}^{j_a}(W'') \vert i \rangle =
\]
\[
\langle f' \vert \tilde{V}_{a}^{j_a} \vert i \rangle + 
{\langle f' \vert \tilde{V}_a^{j_{a}}\vert b \rangle 
\langle b \vert \tilde{V}^{j_a}_a \vert i \rangle  
\over W'' - W_b + i0^+} +
\]
\begin{equation}
\int d[a''']
{\langle f'' \vert 
\tilde{V}_a^{j_{a}}\vert a^{\prime\prime\prime+} \rangle \langle 
a^{\prime\prime\prime+} \vert 
\tilde{V}_a^{j_a} \vert i \rangle 
\over W'' - W'''_{a} + i0^+} 
\label{eq:GY}
\end{equation}
where $\vert b \rangle$ denotes the two body bound states
(we assume here that there is only one corresponding to the deuteron). 
Subtracting (\ref{eq:GY}) from the same equation with $W'' \to W_i$ gives
\[
\langle f' \vert \tilde{T}_{a}^{j_a}(W'') \vert i \rangle =
\langle f' \vert \tilde{T}_{a}^{j_a}(W_i) \vert i \rangle +
\]
\[
\langle f' 
\vert b \rangle \langle b \vert i \rangle [W_b-W_{f'}][W_b - W_i]
\times 
\]
\[     
[{1\over W'' - W_b + i0^+}- {1\over W_i - W_b + i0^+}] + 
\]
\[
\int d[a'''] \langle f' 
\vert \tilde{T}_{a}^{j_a}(W''')\vert a''' \rangle 
\langle a''' \vert \tilde{T}_{a}^{j_a}(W''') \vert i \rangle
\times
\]
\[
[W'''-W_{f'}][W''' - W_i] \times
\]
\begin{equation}     
[{1\over W'' - W''' + i0^+}- {1\over W_i - W''' + i0^+}] 
\label{eq:GZ}
\end{equation} 
which expresses the off-shell transition matrix elements in terms of
the half shell matrix elements using the spectral decomposition.
These methods were used in the original Gl\"ockle, Lee and Coester 
paper and by Gl\"ockle and
Kamada~\cite{gloeckle3}.  We also used this method in our calculations of
in the next section. 

An alternative is to use the first resolvent equation with the definition of 
$\tilde{T}_a^{j_a}(W)$ to shift the value of $W$
\[
\langle  f' | \tilde{T}_{a}^{j_a} (W) 
| i \rangle =
\]
\[ 
{ w_f + w_i \over \sqrt{w_f^2 +K_a^2}+\sqrt{w_i^2 +K_a^2}}
\langle f \vert \bar{T}^{j_a}_a (w_f) \vert i \rangle  +
\]
\[
\int { w_f + w'' \over \sqrt{w_f^2 +K_a^2}+\sqrt{w^{\prime \prime 2} +K_a^2}}
\langle f \vert \bar{T}^{j_a}_a (w_f) \vert a'' \rangle d[a''] \times  
\]
\[
{W_f -W \over
(\sqrt{w_f^2 +K_a^2}+\sqrt{w^{\prime \prime 2} +K_a^2})
(W-W'') }\times
\]
\begin{equation}
\langle a'' \vert \tilde{T}^{j_a}_a (W)\vert i \rangle 
\label{eq:GZA}
\end{equation}
This uses left half shell two-body transition operators to calculate
the off-shell input to the Faddeev-Lovelace equations.  The left 
half-shell two-body transition matrix elements have similar relations to
the non-relativistic left half shell two-body transition transition
matrix elements as the right half off shell transition matrix
elements.  This method can be used to systematically 
study the size of the corrections due to the off shell effects.  

The kernel of the Faddeev Lovelace equations has the form
\[
\langle (ij)(k) \vert (i'j')(k') \rangle 
{\langle (i'j')(k') \vert \tilde{T}_{(i'j')(k')}(W)\vert (i'j')(k') \rangle  
\over W - M_0'+i0^+ } 
\]
Equations (\ref{eq:GXA}), (\ref{eq:GXB}) or (\ref{eq:GXC}) give 
expressions for  $\langle (i'j')(k') \vert \tilde{T}_{(i'j')(k')}(W)
\vert (i'j')(k') \rangle$,
when $W=W'$,  in terms of the non-relativistic 
two-body transition operator. 

The non-relativistic kernel has a similar structure.  The recoupling
coefficients $\langle (ij)(k) \vert (i'j')(k') \rangle$, which relate
different orders of coupling of Poincar\'e irreducible representations
are replaced by coefficients that relate different orders of coupling $SU(2)$
irreducible representations.  The matrix elements $\langle (i'j')(k')
\vert \tilde{T}_{(i'j')(k')}(W)\vert (i'j')(k')\rangle$ 
are replaced by off-shell two
body transition matrix elements, and the energy denominator ${1/( W -
M_0')}$ is replaced by the corresponding non-relativistic energy
denominator.

The energy denominator and transition matrix elements in the relativistic
and non-relativistic case are related by kinematic factors when 
$\tilde{T}$ is on the right half shell; this is precisely the
point where the energy denominator has an integrable singularity.  
This suggests that the off shell contributions for $\tilde{T}$
might be suppressed in the dynamical equations.  
 
Relativistic effects also appear in the permutation operators 
$\langle (ij)(k) \vert (i'j')(k')\rangle$.  The difference between the 
relativistic and non-relativistic coefficients can be easily seen 
when the recoupling  coefficients are computed using the
Balian-Brezin method \cite{jean}\cite{balian}. 

The non-relativistic overlap matrix that relates the 
(12)(3) coupling scheme to the (23)(1) coupling scheme
is:
\[
\langle {\bf P}, j, \mu, (23)(1) 
\vert  
{\bf P}' j' \mu' (12)(3) \rangle =
\] 
\[
\delta ({\bf P} - {\bf P}') \delta (E-E') 
{\delta_{jj'} \delta_{\mu \mu'}  
\over 2j+1} \times
\]
\[ 
\sum_{\mu''=-j}^j \sum
{8 \pi^2 (m_1+m_2)(m_1+m_3) \over
m_1 m_2 m_3 k_{23} k_{12} K_{23} K_{12}}
\times
\]
\[
Y^*_{L_{23} m_L} (\hat{\bf K}_{23})
Y^*_{l_{23} m_l} (\hat{\bf k}_{23})
Y_{L_{12}' m_L'} (\hat{\bf K}_{12})
Y_{l_{12}' m_l'} (\hat{\bf k}_{12}) \times
\]
\[
\langle j, \mu'' \vert L_{23}, m_L, S_{23}, m_S \rangle
\langle S_{23}, m_S \vert s_1, m_1, j_{23}, \mu_{23} \rangle \times
\]
\[
\langle j_{23}, \mu_{23} \vert l_{23}, m_l, s_{23}, m_{23} \rangle
\langle S_{23}, m_{23} \vert s_2, m_2, s_3, m_3 \rangle \times
\]
\[
\langle S_{12} m_{12} \vert s_1, m_1, s_2, m_2 \rangle
\langle j_{12} \mu_{12} \vert l_{12}', m_l', s_{12}, m_{12} \rangle
\times 
\]
\begin{equation}
\langle S_{12}' m_S' \vert s_3, m_3, j_{12}, \mu_{12} \rangle
\langle J, \mu'' \vert L_{12}', m_L', S_{12}', m_S' \rangle
\label{eq:GAA}
\end{equation}
where the angular variables can be computed for {\it any} convenient 
set of vectors of 
${\bf k}_{12}$, ${\bf K}_{12}$, ${\bf k}_{23}$, ${\bf K}_{23}$ that have the 
correct values of $k_{12}^2$, $k_{23}^2$, $K_{12}^2$ and $K_{12}^2$.

An identical computation can be done in the relativistic case with 
the following result: 
\[
\langle {\bf P}, j, \mu, (23)(1) 
\vert  
{\bf P}' j' \mu' (12)(3) \rangle =
\] 
\[
\delta ({\bf P} - {\bf P}') \delta (M^{(0)}-M^{(0)\prime}) 
{\delta_{jj'} \delta_{\mu \mu'}  
\over 2j+1} 
\sum_{\mu''=-j}^j \sum
\times
\]
\[
{8 \pi^2 \over  k_{12} k_{23} K_{12} K_{23}} 
\left [
{(\omega_1 (k_{12})+ \omega_2 (k_{12}))^3 
\over 
\omega_1 (k_{12}) \omega_2 (k_{12}) \omega_3 (K_{12}) \omega_{12} (K_{12})}
\right ]^{1/2} \times 
\]
\[
\left [ {(\omega_2 (k_{23})+ \omega_3 (k_{23}))^3 \over
\omega_2 (k_{23}) \omega_3 (k_{23}) \omega_1 (K_{23}) 
\omega_{23} (K_{23})}\right ]^{1/2}
\times
\]
\[
Y^*_{L_{23} m_L} (\hat{\bf K}_{23})
Y^*_{l_{23} m_l} (\hat{\bf k}_{23})
Y_{L_{12}' m_L'} (\hat{\bf K}_{12})
Y_{l_{12}' m_l'} (\hat{\bf k}_{12}) \times
\]
\[
\langle J, \mu'' \vert L, m_L, S, m_S \rangle
\langle S, m_S \vert s_1, m_1, j_{23}, \mu_{23} \rangle \times
\]
\[
\langle j_{23}, \mu_{23} \vert l, m_l, s_{23}, m_{23} \rangle
\langle S_{23}, m_{23} \vert s_2, m_2, s_3, m_3 \rangle \times
\]
\[
D^{s_1}_{m_1 \bar{m}_{1}} [R_w (L({\bf K}_{12}),{\bf k}_{12}') ] 
D^{s_3}_{m_3 \bar{m}_{3}} [R_w^{-1} (L({\bf K}_{23}),{\bf k}_{32}) ] 
\times
\]
\[
D^{s_2}_{m_2 \bar{m}_{2}} [R_w^{-1} (L({\bf K}_{23}),{\bf k}_{23}) 
R_w (L({\bf K}_{12}),{\bf k}_{21}) ] \times
\]
\[
\langle S_{12} m_{12} \vert s_1, \bar{m}_1, s_2, \bar{m}_2 \rangle
\langle j_{12} \mu_{12} \vert l', m_l', s_{12}, m_{12} \rangle \times
\]
\begin{equation}
\langle S' m_S' \vert s_3, \bar{m}_3, j_{12}, \bar{\mu}_{12} \rangle
\langle J, \mu'' \vert L', m_L', S', m_S' \rangle
\label{eq:GAFA}
\end{equation}

Comparison of the two coefficients shows three
differences; 
\begin{itemize} 
\item [1.] The energy conserving 
delta function $\delta (E-E')$ in (\ref{eq:GAA}) 
is replaced by 
a delta function in the invariant mass 
$\delta (M^{(0)}-M^{(0)\prime})$. 

\item [2.] The second modification is that 
in the relativistic expression the coefficient 
\begin{equation}   
{8 \pi^2 (m_1+m_2)(m_2+m_3) \over
m_1 m_2 m_3 }
\label{eq:GAC}
\end{equation}
\item[] is replaced by 

\[
{8 \pi^2 \over  k_{12} k_{23} K_{12} K_{23}} 
\left [
{(\omega_1 (k_{12})+ \omega_2 (k_{12}))^3 
\over 
\omega_1 (k_{12}) \omega_2 (k_{12}) \omega_3 (K_{12}) \omega_{12} (K_{12})}
\right ]^{1/2} \times 
\]
\begin{equation}
\left [ {(\omega_2 (k_{23})+ \omega_3 (k_{23}))^3 \over
\omega_2 (k_{23}) \omega_3 (k_{23}) \omega_1 (K_{23}) 
\omega_{23} (K_{23})}\right ]^{1/2}
\label{eq:GAD}
\end{equation}

\item [] which agrees with the non-relativistic expression in the limit that 
all of the relative momenta vanish.

\item [3.] The third modification to the recoupling coefficients involves the
addition of momentum-dependent rotations that appear between the
$SU(2)$ Clebsch-Gordan coefficients. 
\end{itemize} 
This choice of recoupling coefficient 
assumes that the spin in the relativistic problem is
the canonical spin. 
In the relativistic expression $R_w(\Lambda ,{\bf p})$ is the Wigner rotation  
\begin{equation}
R_w (\Lambda ,{\bf p} ) := L^{-1} (\Lambda p) \Lambda L({\bf p})
\label{eq:GAG}
\end{equation}
where $L({\bf p})$ is the rotationless Lorentz transformation boosts a
particle at rest to momentum ${\bf p}$.  The subscript 
${\bf K}_{23}$ indicates that the starting four vector is
\[
(\omega_{23} , {\bf K}_{23}) ) =
\]
\begin{equation}
\left ( \sqrt{(\sqrt{m_2^2 + 
{ k_{23}}^2}+ \sqrt{m_3^2 + { k_{23}}^2})^2 +
{ K}_{23}^2},  
-{\bf K}_{23} \right ). 
\label{eq:GAH}
\end{equation}
As with the multiplicative factors, when all of the momenta vanish 
these rotations become the identity, and the expression reduces to 
the non-relativistic expression. 

In the Balian-Brezin form the transition from the non-relativistic 
to relativistic recoupling coefficients involves including some
simple momentum dependent factors. 

The form of the relativistic recoupling coefficients will depend on 
the specific representation of the relativistic dynamics.  The coefficients 
given above are in an ``instant-form'' representation.

\section{Results and Discussion}
\label{sec:results}

In this section consider the contributions of various terms to the 
kernel of the relativistic Faddeev Lovelace equations.

The most complex modification of the non-relativistic Faddeev Lovelace 
equations involves the treatment of the off-energy shell kernel.  As was 
mentioned in the previous section, in order to compute this from a
non-relativistic two body transition operator fit to data, it was 
necessary either to utilize a full spectral expansion or to solve an 
integral equation.  Both of these procedures involve a significant effort
beyond that required to solve the non-relativistic equation.
On the other hand we showed that the half-on-shell kernel could be 
calculated exactly in terms of the non-relativistic transition 
matrix elements by introducing only multiplicative kinematic factors.
In addition, the on shell value was closest to scattering singularity 
in the matrix elements of the three body free resolvent.
  
The first test that we perform is to investigate the impact of simply
including the kinematic factors that are needed to construct the
half-on-shell kernel.  These approximations are given by equations
(\ref{eq:GXA}), (\ref{eq:GXB}), and (\ref{eq:GXC}).

To set these approximations we consider a simple model without spin.
The non-relativistic dynamics is given by the Malfliet-Tjon V
interaction~\cite{malfliet}, which has a long range attractive part
and a short range repulsive core.  Using this model we compare
\begin{itemize}
\item[1.] The exact calculation of off shell matrix elements of 
$\tilde{T}(W)$ computed using the spectral expansion~(\ref{eq:GZ}).

\item[2.] These approximations of $\tilde{T} (W)$ given by equations
  (\ref{eq:GXA}), (\ref{eq:GXB}), and (\ref{eq:GXC}), and by keeping
  only the first term in Eq.~(\ref{eq:GZ}), ignoring the dispersive
  contributions.

\item[3.] The non-relativistic off shell transition operator.
\end{itemize}

The following kinematic variables were varied in these comparisons:
\begin{enumerate}
\item $W = 3m + E_{NN}$ (three-body energy);
\item ${\bf K}_{ij}$ (spectator momentum);
\item ${\bf k}$ (initial two-body momentum);  
\item ${\bf k}' = {\bf k}$ (initial=final two-body momentum).
\end{enumerate}
We also considered all three methods of using the non-relativistic  
two-body dynamics as input.  

The results of these calculations are shown in figures 3 - 14.  The
plots compare the difference between the non-relativistic and exact
results divided by the exact result.  This is done for each method.
Figures 3-6 use the CPS method, figures 7-10 plot the same quantities 
for the GLC method, and figures 11-14 plot the same quantities for 
the GK method. 
The first plot in each set varies the off shell energy for initial
and final values of $k_{ij}$ equal and for a fixed value of
$K_{ij}$.  The next plot fixes the energy $K_{ij}$ and varies the
initial and final values of $k_{ij}$.  The third plot varies the
final $k_{ij}$ keeping everything else fixed and the fourth plot
varies $K_{ij}$ holding everything else constant.

In the CPS and GLC cases the error in the approximations
(\ref{eq:GXA}) or (\ref{eq:GXB}) to the exact off shell matrix
elements is no more that a few tenths of a percent in all four cases.
They are significant improvements over the straight non-relativistic
result.  In the GK case the errors can be as much as 5\% percent in
some kinematic regions, although the errors are typically about twice
the size of the errors using the other two methods.

These calculation suggest that the bulk of the relativistic 
effects in the transition operators in the Faddeev Lovelace kernel 
comes from the kinematic factors that appear in      
(\ref{eq:GXA}), (\ref{eq:GXB}) or (\ref{eq:GXC})

The presence of the free resolvent in the Faddeev Lovelace kernel 
only serves to improve the approximation to the full kernel, 
since it is largest on the half shell. 

%[ Question - what units are being used.  Is the
%energy in $t_{nr}$ kept on shell in these calculation? ]

\begin{figure}
\begin{center}
\rotatebox{270}{\resizebox{2.9in}{!}{
\includegraphics{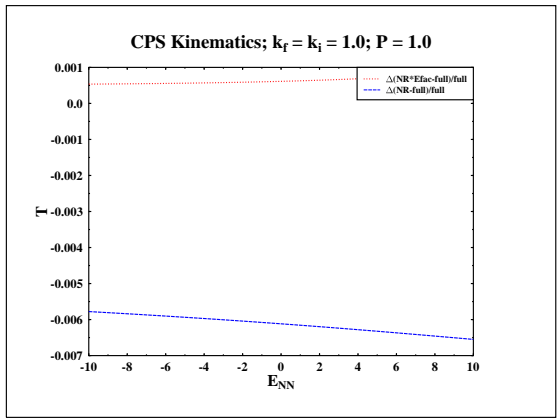}}
}
\end{center}
\label{fig:CPSE}
\caption{Fractional deviation of approximate $T$ matrix compared to exact
  evaluation of Eq.~(\ref{eq:GZ}) using the CPS method, as a function of
  the 2-body energy (MeV) in the 3-body CM frame.  Dotted line uses
  approximate value described in the text.  Dashed line uses the
  nonrelativistic value.}

\end{figure}

\begin{figure}
\begin{center}
  \rotatebox{270}{\resizebox{2.9in}{!}{
      \includegraphics{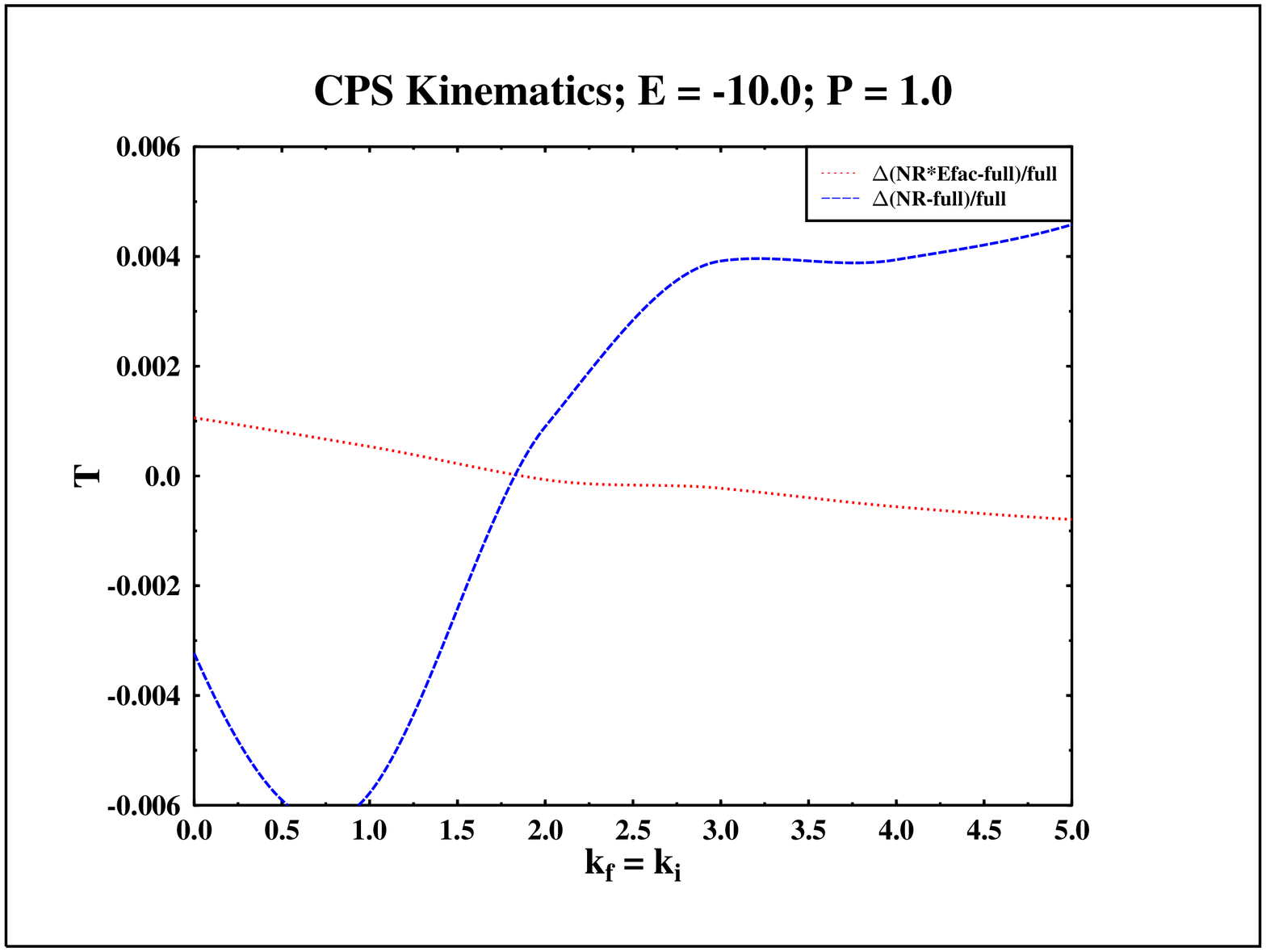}} }
\end{center}
\label{fig:CPSK1}
\caption{Fractional deviation of approximate $T$ matrix compared to exact
  evaluation of Eq.~(\ref{eq:GZ}) using the CPS method, as a function of
  the 2-body relative momentum (fm$^{-1}$), initial and final momenta
  kept equal.  Dotted line uses approximate value described in the
  text.  Dashed line uses the nonrelativistic value.}

\end{figure}

\begin{figure}
\begin{center}
\rotatebox{270}{\resizebox{2.9in}{!}{
\includegraphics{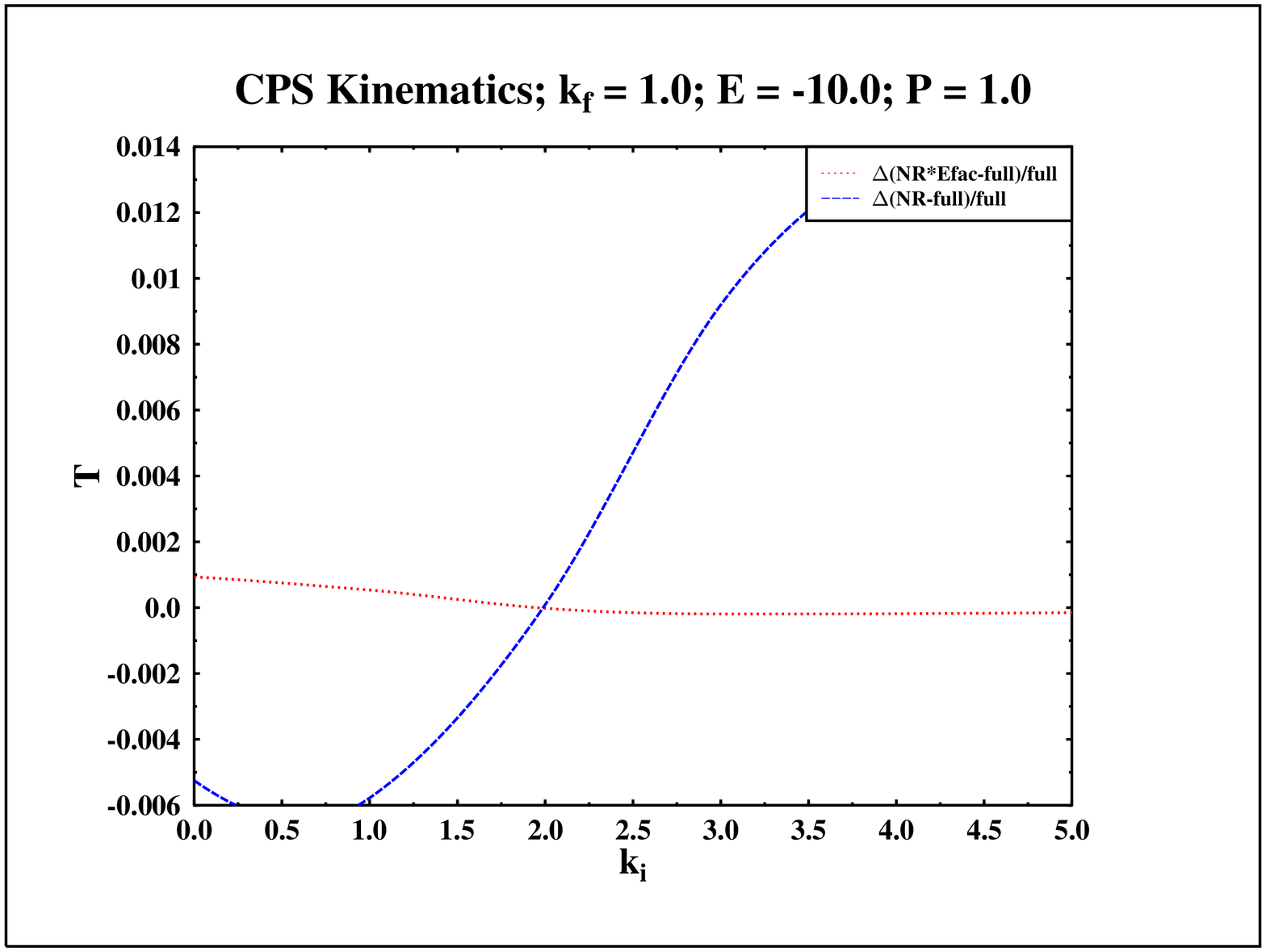}}
}
\end{center}
\label{fig:CPSK2}
\caption{Fractional deviation of approximate $T$ matrix compared to exact
  evaluation of Eq.~(\ref{eq:GZ}) using the CPS method, as a function of
  the 2-body initial relative momentum (fm$^{-1}$).  Dotted line uses
  approximate value described in the text.  Dashed line uses the
  nonrelativistic value.}

\end{figure}

\begin{figure}
\begin{center}
\rotatebox{270}{\resizebox{2.9in}{!}{
\includegraphics{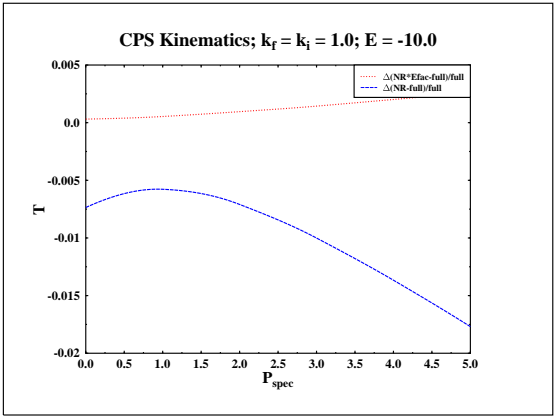}}
}
\end{center}
\label{fig:CPSP}
\caption{Fractional deviation of approximate $T$ matrix compared to exact
  evaluation of Eq.~(\ref{eq:GZ}) using the CPS method, as a function of
  the spectator momentum (fm$^{-1}$).  Dotted line uses approximate
  value described in the text.  Dashed line uses the nonrelativistic
  value.}

\end{figure}

\begin{figure}
\begin{center}
\rotatebox{270}{\resizebox{2.9in}{!}{
\includegraphics{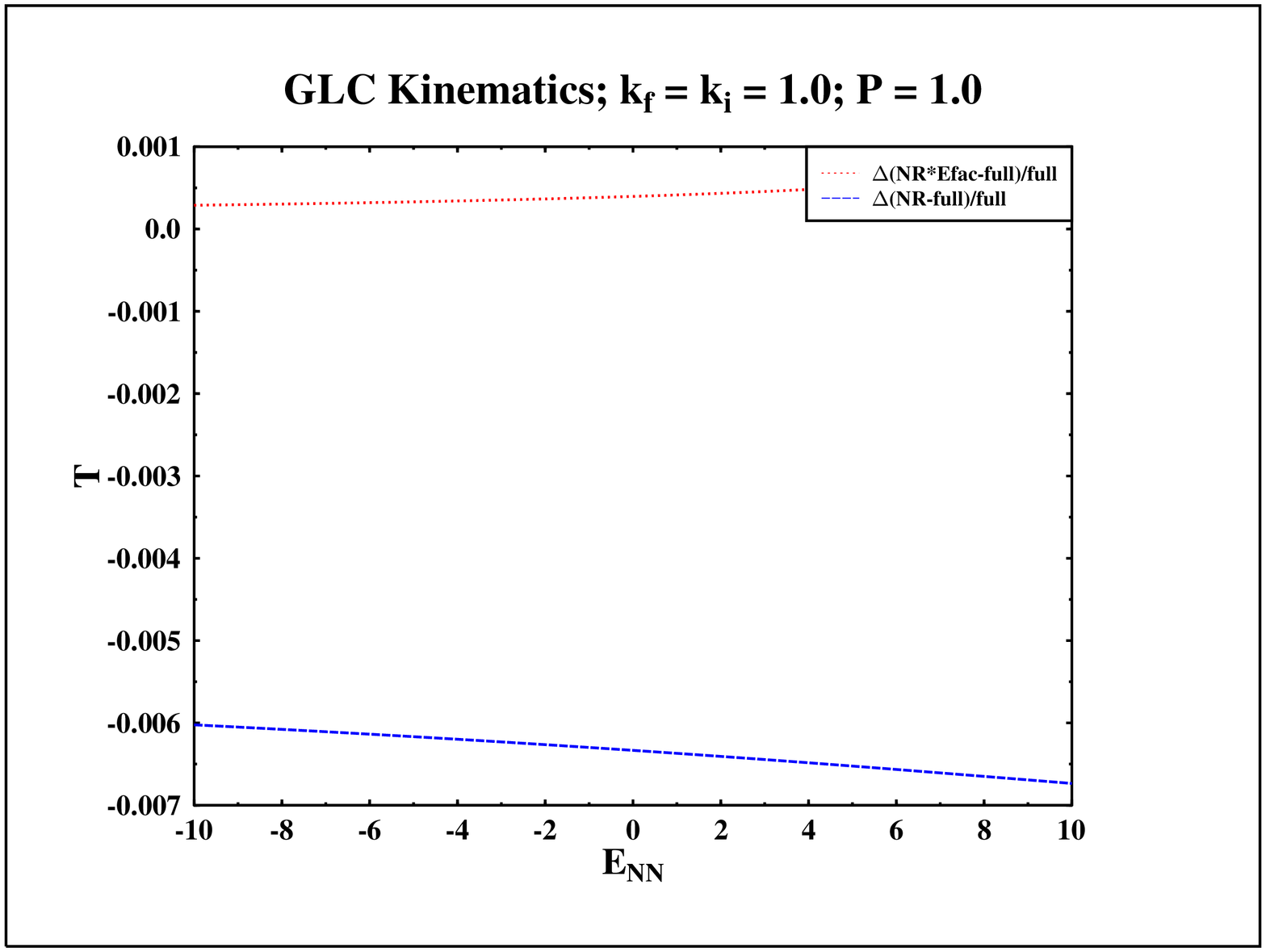}}
}
\end{center}
\caption{Same as Fig.~3 except using the GLC method.}

\end{figure}

\begin{figure}
\begin{center}
\rotatebox{270}{\resizebox{2.9in}{!}{
\includegraphics{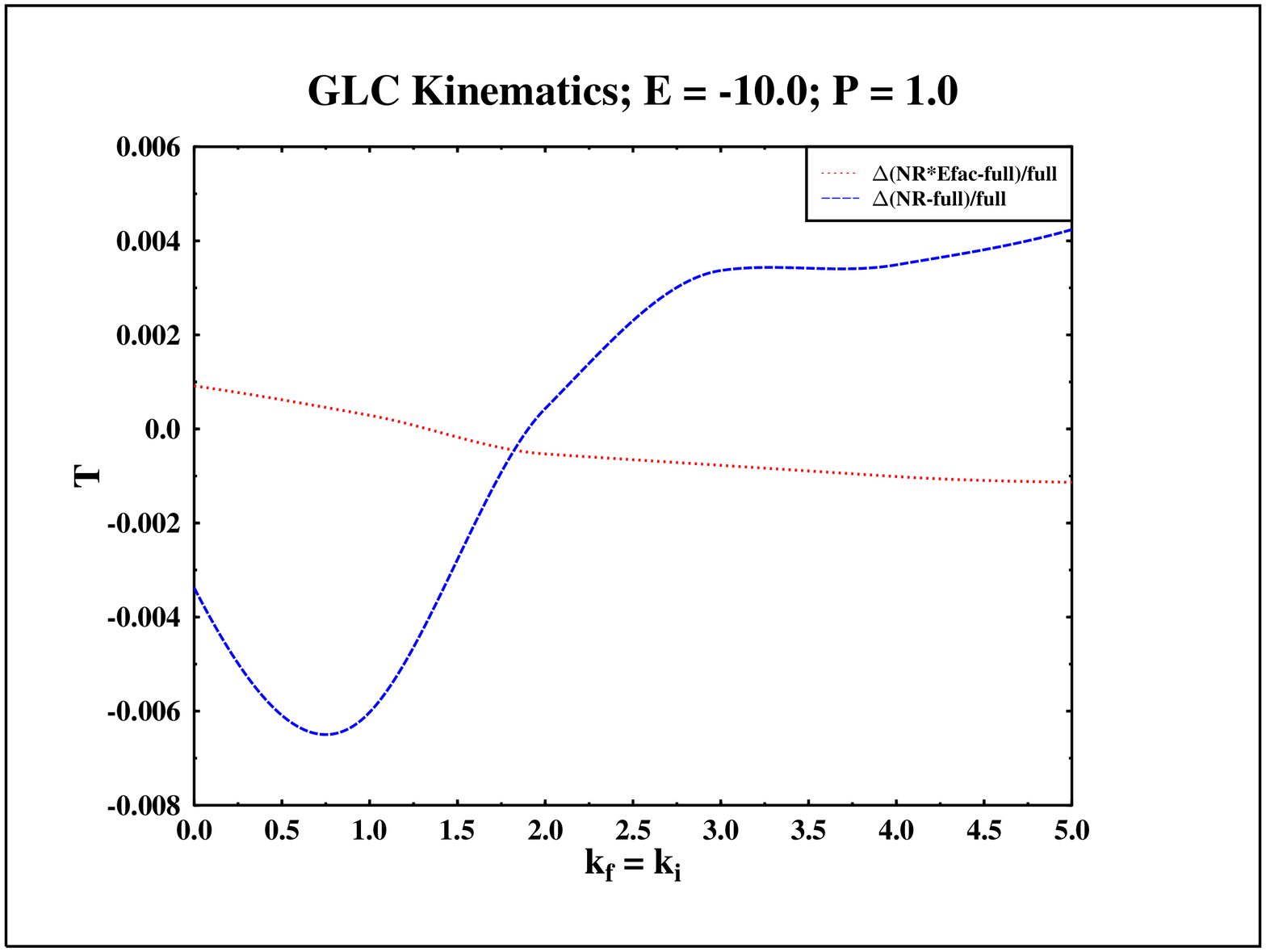}}
}
\end{center}
\caption{Same as Fig.~4 except using the GLC method.}

\end{figure}

\begin{figure}
\begin{center}
\rotatebox{270}{\resizebox{2.9in}{!}{
\includegraphics{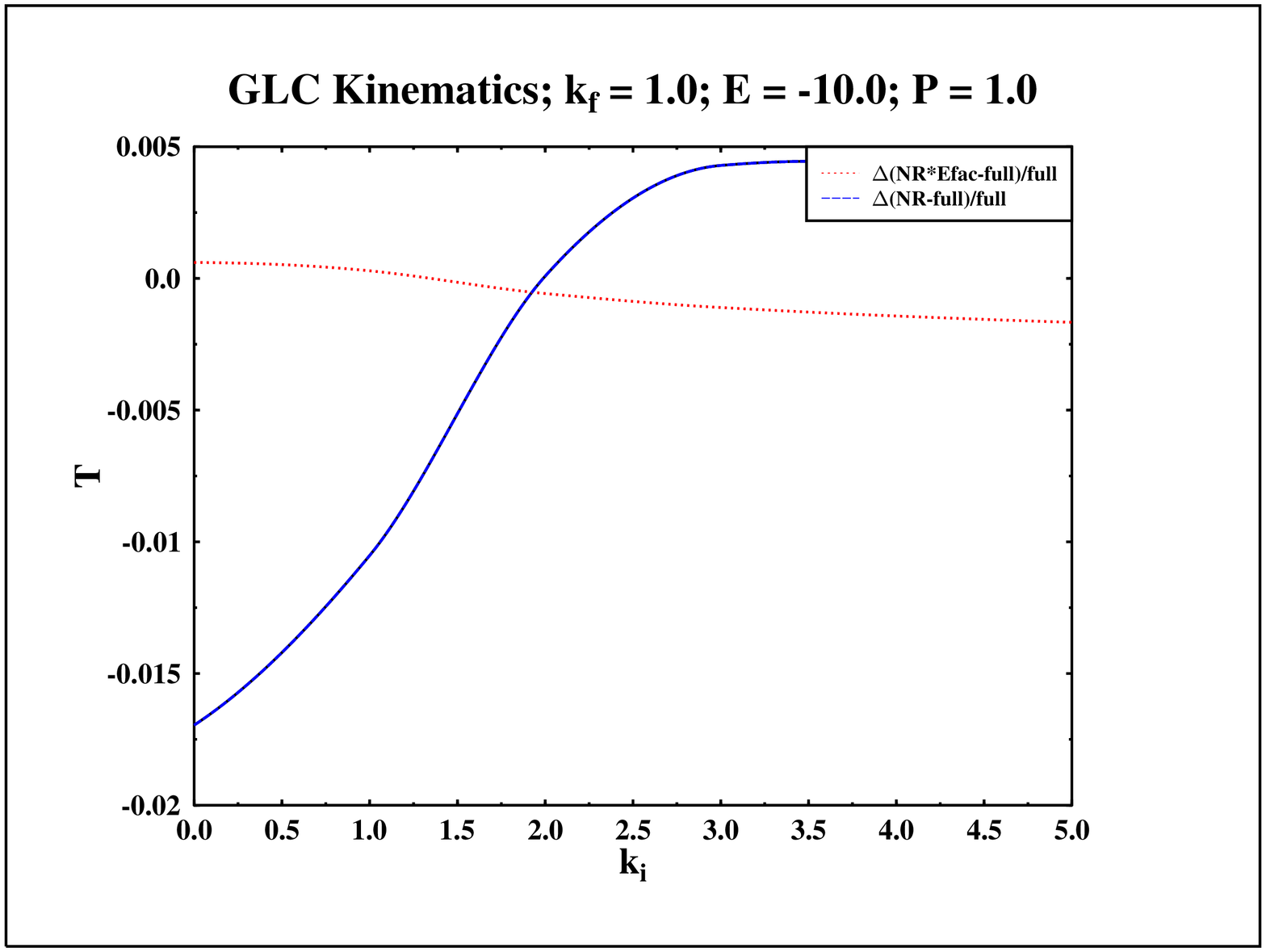}}
}
\end{center}
\caption{Same as Fig.~5 except using the GLC method.}

\end{figure}

\begin{figure}
\begin{center}
\rotatebox{270}{\resizebox{2.9in}{!}{
\includegraphics{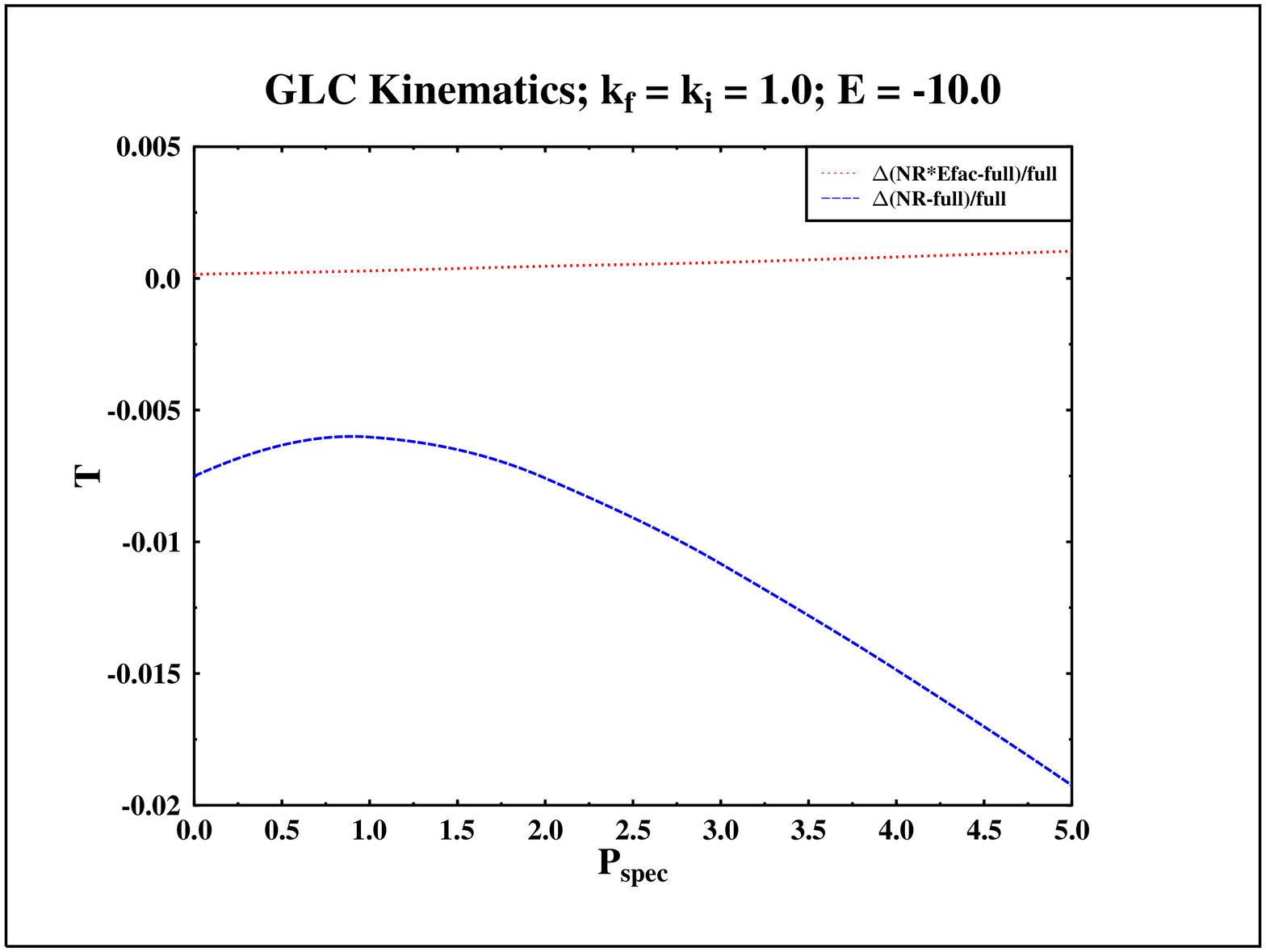}}
}
\end{center}
\caption{Same as Fig.~6 except using the GLC method.}

\end{figure}

\begin{figure}
\begin{center}
\rotatebox{270}{\resizebox{2.9in}{!}{
\includegraphics{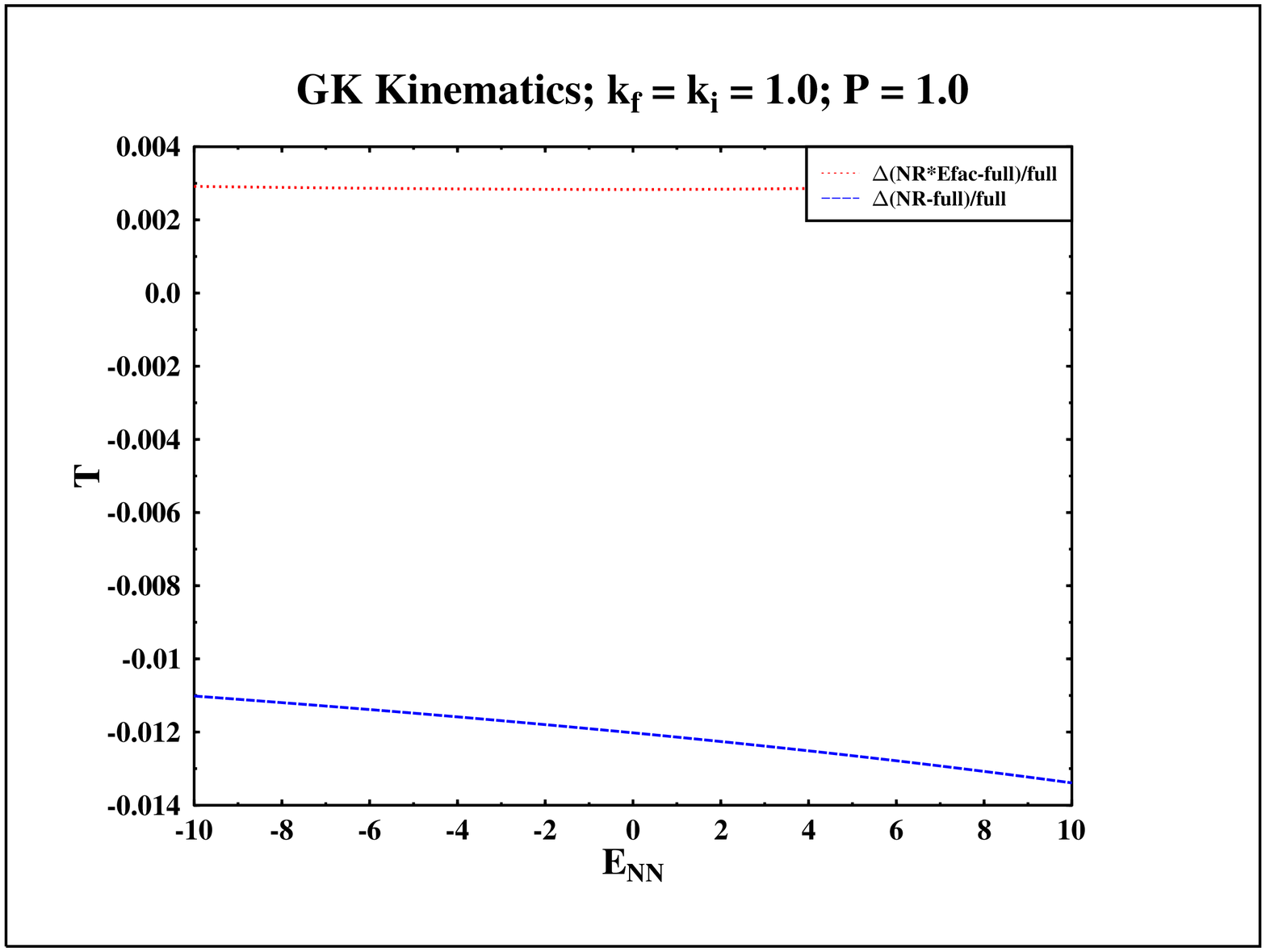}}
}
\end{center}
\caption{Same as Fig.~3 except using the GK method.}

\end{figure}

\begin{figure}
\begin{center}
\rotatebox{270}{\resizebox{2.9in}{!}{
\includegraphics{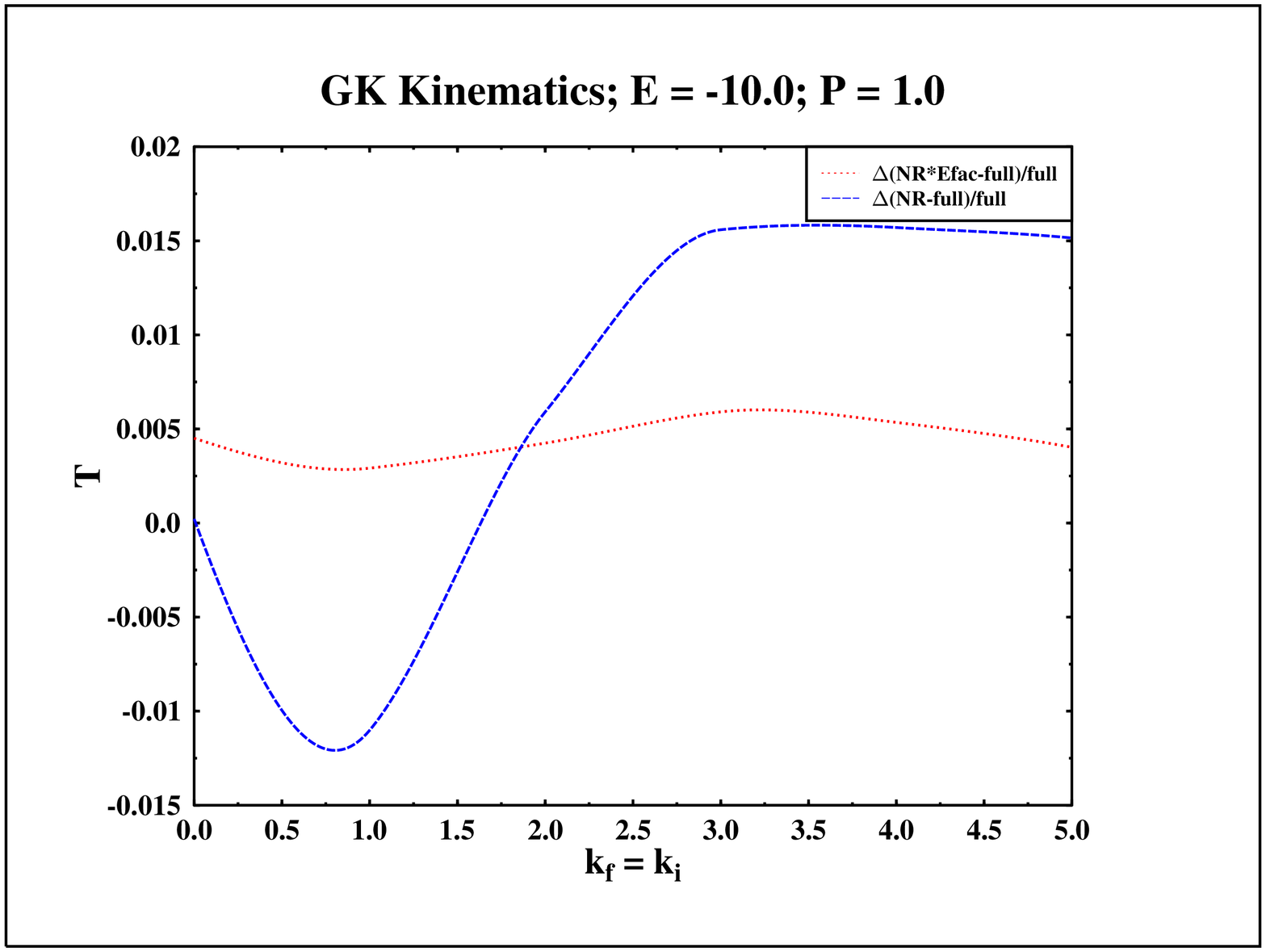}}
}
\end{center}
\caption{Same as Fig.~4 except using the GK method.}

\end{figure}

\begin{figure}
\begin{center}
\rotatebox{270}{\resizebox{2.9in}{!}{
\includegraphics{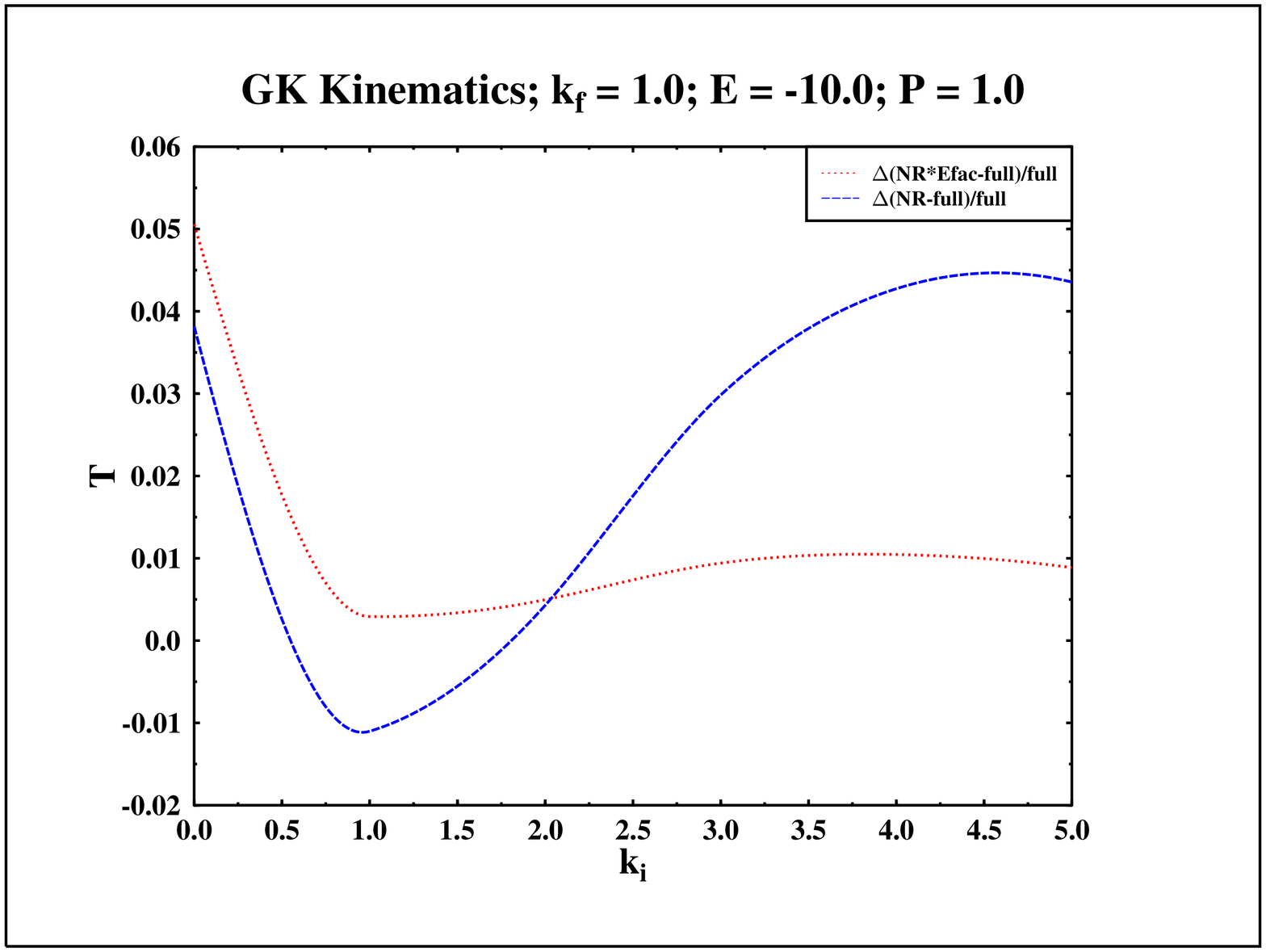}}
}
\end{center}
\caption{Same as Fig.~5 except using the GK method.}

\end{figure}

\begin{figure}
\begin{center}
\rotatebox{270}{\resizebox{2.9in}{!}{
\includegraphics{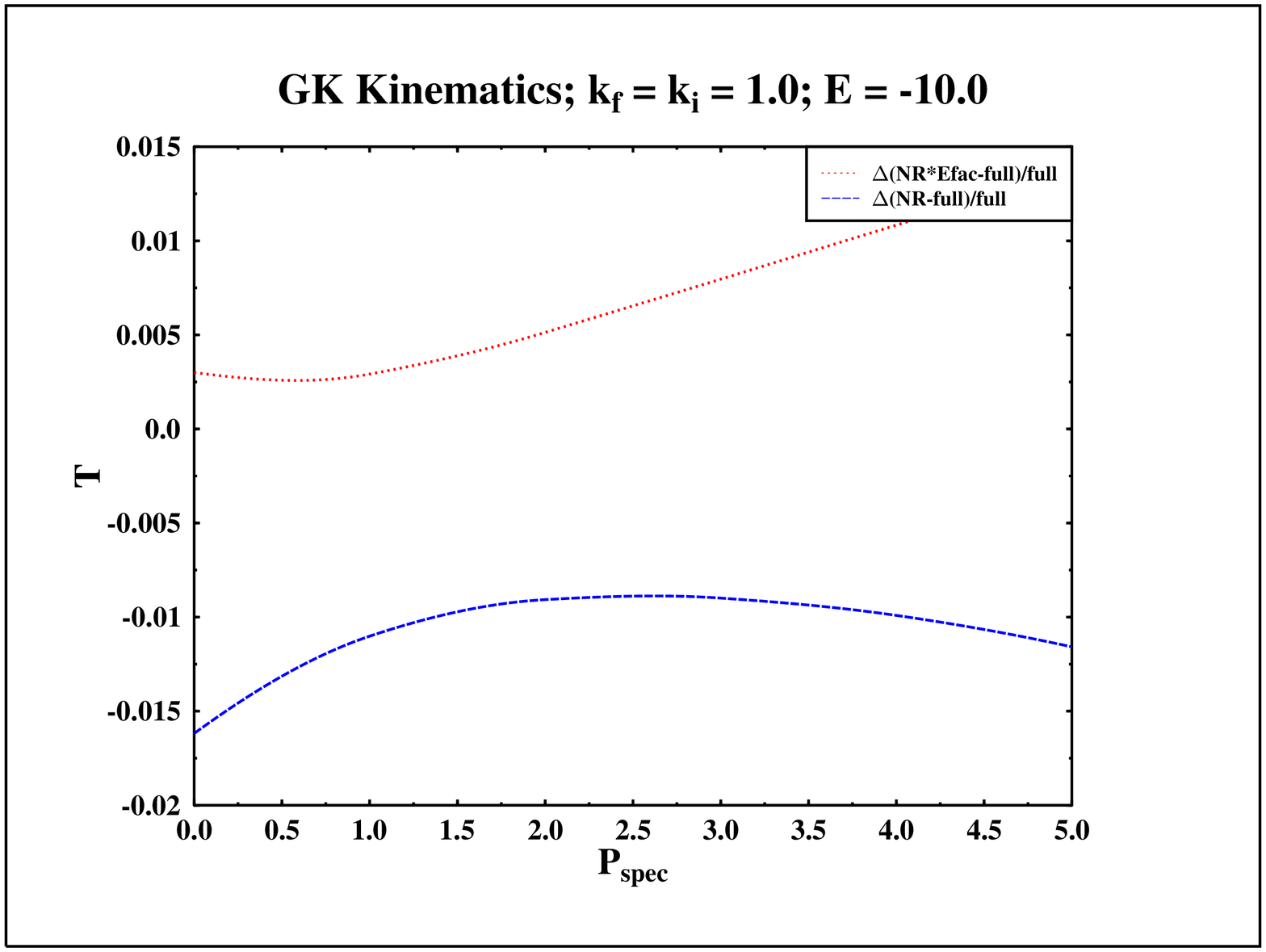}}
}
\end{center}
\caption{Same as Fig.~6 except using the GK method.}

\end{figure}

The other factor that appears in a full relativistic three 
body calculation are the three body recoupling coefficients.  
These differ from the non-relativistic coefficients 
by kinematic factors and the presence of momentum dependent 
spin rotation functions.  In many relativistic calculations
the overall contribution of the spin effects have proved to be 
unimportant, however most of these applications have involved 
bound states. 

\begin{figure}
\begin{center}
\rotatebox{270}{\resizebox{2.9in}{!}{
\includegraphics{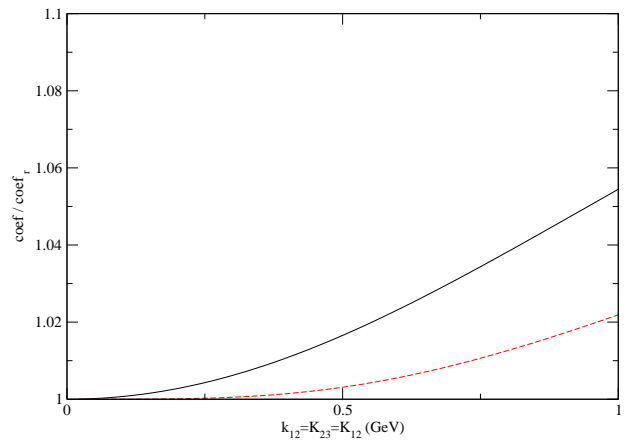}}
}
\end{center}
\label{fig:recoupling}
\caption{Ratio of the recoupling 
coefficients.}
%}
\end{figure}

It is a simple matter to compute relativistic and nonrelativistic
recoupling coefficients.  To compare them the, delta function in
either invariant mass or non-relativistic internal energy is replaced
by a delta function in the momentum followed by the appropriate
Jacobian factors.  The comparison can then be made between the
coefficients of the delta functions in the two cases.
Figure~\ref{fig:recoupling} is typical of the behavior of the
recoupling coefficient.  The curves are computed for $k_{12}=
K_{23}=K_{12}$ for a representative set of spin quantum numbers.  The
solid curve is the ratio of the non-relativistic to the relativistic
recoupling coefficient when they are made to have the same delta
function.  The dotted curve shows the ratio of relativistic
coefficient with the rotation functions turned off to the full
relativistic recoupling coefficient.

The curves show that the non-relativistic curve is about 2\% 
higher than the relativistic curve at 0.5~GeV.  It grows to just over 
5\% at 1~GeV.    Including the correct multiplicative kinematic factors
without including spin rotations leads to considerable improvement 
at 0.5~GeV; and it grows to just over 2\% at 1~GeV.
This is consistent with previous calculations that suggest the 
the recoupling coefficients have small effects. 

With the Balian-Brezin method the inclusion of both the spin factors and 
kinematic factors involve minor modifications to an 
existing non-relativistic program. 

%% {\it
%% Note: plots and discussions go here.  My ps files are in disarray
%% right now.  It's been long enough that I don't remember how they were
%% generated.  What I'd like to have is sweeps across kinematic variables
%% using the potential that generates the highest momentum components
%% (MT-V?), with the NR and NR-sqrt results plotted as fractional
%% differences with the full results.
%% }

Our conclusion is that negligible errors will be made if the
relativistic Faddeev-Lovelace kernel is approximated by replacing the
transition matrix elements of $\tilde{T}$ by the expressions in
(\ref{eq:GXA}),(\ref{eq:GXB}), and (\ref{eq:GXC}).   Care must be
used to match the method used to the fitting procedure used to 
determine the non-relativistic two body interaction.  The 
recoupling coefficients are not needed for calculations dominated by 
lower momenta; and they lead $5\%$ effects at 1~GeV.
The inclusion of the correct kinematic factors reduces errors
considerably.
 
\section{Comparison to Approaches Based On Quasipotential Theory}
\label{sec:field}

The development and the results discussed above utilize Wigner's
formulation of relativistic quantum mechanics, using a direct
construction of a unitary representation of the Poincar\'e group on
the two-nucleon Hilbert space~\cite{wigner}.  The two-nucleon
interactions have specific connections both to the Schr\"odinger
equation and to the three-nucleon problem.  One can also approach
these problems within the framework of a quasipotential theory.  The
first results for three-nucleon binding were obtained by Stadler and
Gross~\cite{stadler}, using the so-called spectator
approximation~\cite{gross1,gross2,gross3}.  Within this framework they
can obtain the observed triton binding while maintaining a fit to the
observed nucleon-nucleon phase shifts and the deuteron binding.  When
they compare the results of their ``full'' calculation to a
nonrelativistic approximation, they find that most of the new effects
come from the inclusion of negative-energy states.  These would
correspond approximately to three-nucleon interactions in Poincar\'e invariant
quantum theory.  They also find that so-called ``boost effects'' in
their framework are small and give a repulsive contribution to the
triton binding.  Those results are consistent with 
the results obtained in Poincar\'e invariant quantum theory.

An earlier work by Sammarruca and Machleidt~\cite{sammarruca} examined
the effect of kinematic factors within frameworks based on field
theories.  They made use of an earlier relation obtained by Brown,
Jackson and Kuo~\cite{bjk}, who used quasipotential methods to argue
that the non-relativistic interaction should include non-localities.
The Brown, Jackson, Kuo method leads to a modified interaction,
however unlike the methods discussed above the new interaction must be
refit to the phase shift data.  This method leads to the following
relation between their minimal relativistic and nonrelativistic $t$
matrices:
\begin{equation}
  \label{eq:1s50}
  t_{R}({\bf k}' \, {\bf k}) =
  \sqrt{\omega_k'\over m} 
  t_{NR}({\bf k}' \, {\bf k}) \sqrt{\omega_k\over m}.
\end{equation}
%The origin of these factors appears to come from these authors' use of
%covariant normalization of density of states:
%\begin{equation}
%  \label{eq:1s75}
%  d{\bf k} \to {m\over \omega_k} d{\bf k}.
%\end{equation}
Sammarruca and Machleidt then use this modified two-nucleon
interaction within their relativistic framework and get a net {\it
  attractive} correction to $^3$H binding when using this factor.
Again, this is understandable at least naively when considering that
the $t$ matrices receive essentially the opposite factor from those
discussed above [Eqs.~\ref{eq:GXA}-\ref{eq:GXC}].  However, both the
Brown, Jackson and Kuo method and the CGL method in principle require
a refitting of the phase shifts to get an exact phase equivalence.
Thus we cannot make a direct comparison to these methods.
%
% this
%does not explain the difference between this approach and that of
%Stadler and Gross~\cite{stadler}, who obtain a net {\it repulsion} on
%the basis of internal kinematics.  Without further details of both
%works, it is difficult to pursue this comparison further.

\section{Conclusion}
\label{sec:conclusion}

We have explored here the quantitative impact of the requirement of
relativistic invariance with the framework of Poincar\'e invariant
quantum mechanics.  In the range of models discussed here, the
dominant effects are manifested in kinematic terms that can easily be
incorporated into a Schr\"odinger-based momentum-space calculation,
namely, multiplicative factors of the $E/m$ form, and energy
denominators that employ the relativistic energy-momentum relation.
There are additional small effects that come from relativistic
corrections to the recoupling coefficients.  These can be minimized by
including the correct relativistic kinematic factors.  A correct
treatment of these terms involves only minimal modifications of the
non-relativistic recoupling coefficients.  Quantitatively the effects
of all of these corrections are small, and contribute repulsive
corrections to three-nucleon binding.  This means that the resolution
of the discrepancy between Poincar\'e invariant three-nucleon
calculations and the experimentally observed binding energies must
come from three-nucleon forces.  The quantitative relativistic effects
grow with internal momenta; non-relativistic three-nucleon scattering
calculations at energies of several hundred MeV may well require
corrections beyond the simple square-root factors.

%Thus, within the 
%Poincar\'e invariant quantum mechanics 
%the discrepancy between the observed three-nucleon binding energy and
%that predicted on the basis of two-nucleon direct interactions alone
%must lie in an additional direct three-nucleon interaction.

\begin{acknowledgments}
This work supported in part by the Office of Science of the U.S. 
Department of Energy, under contract DE-FG02-86ER40286.
\end{acknowledgments}

%\bibliography{3BMS-3}

\begin{thebibliography}{26}
\expandafter\ifx\csname natexlab\endcsname\relax\def\natexlab#1{#1}\fi
\expandafter\ifx\csname bibnamefont\endcsname\relax
  \def\bibnamefont#1{#1}\fi
\expandafter\ifx\csname bibfnamefont\endcsname\relax
  \def\bibfnamefont#1{#1}\fi
\expandafter\ifx\csname citenamefont\endcsname\relax
  \def\citenamefont#1{#1}\fi
\expandafter\ifx\csname url\endcsname\relax
  \def\url#1{\texttt{#1}}\fi
\expandafter\ifx\csname urlprefix\endcsname\relax\def\urlprefix{URL }\fi
\providecommand{\bibinfo}[2]{#2}
\providecommand{\eprint}[2][]{\url{#2}}

\bibitem[{\citenamefont{Wigner}(1939)}]{wigner}
\bibinfo{author}{\bibfnamefont{E.~P.} \bibnamefont{Wigner}},
  \bibinfo{journal}{Ann.\ Math.} \textbf{\bibinfo{volume}{40}},
  \bibinfo{pages}{149} (\bibinfo{year}{1939}).

\bibitem[{\citenamefont{Bakamjian and Thomas}(1953)}]{bt}
\bibinfo{author}{\bibfnamefont{B.}~\bibnamefont{Bakamjian}} \bibnamefont{and}
  \bibinfo{author}{\bibfnamefont{L.~H.} \bibnamefont{Thomas}},
  \bibinfo{journal}{Phys.\ Rev.} \textbf{\bibinfo{volume}{92}},
  \bibinfo{pages}{1300} (\bibinfo{year}{1953}).

\bibitem[{\citenamefont{Keister and Polyzou}(1991)}]{kpreview}
\bibinfo{author}{\bibfnamefont{B.~D.} \bibnamefont{Keister}} \bibnamefont{and}
  \bibinfo{author}{\bibfnamefont{W.~N.} \bibnamefont{Polyzou}},
  \bibinfo{journal}{Adv.\ Nucl.\ Phys.} \textbf{\bibinfo{volume}{20}},
  \bibinfo{pages}{225} (\bibinfo{year}{1991}).

\bibitem[{\citenamefont{Kamada and Gl\"ockle}(1998)}]{gloeckle1}
\bibinfo{author}{\bibfnamefont{H.}~\bibnamefont{Kamada}} \bibnamefont{and}
  \bibinfo{author}{\bibfnamefont{W.}~\bibnamefont{Gl\"ockle}},
  \bibinfo{journal}{Phys.\ Rev.\ Lett.} \textbf{\bibinfo{volume}{80}},
  \bibinfo{pages}{2547} (\bibinfo{year}{1998}).

\bibitem[{\citenamefont{Kamada et~al.}(2002)\citenamefont{Kamada, Gl\"ockle,
  Golak, and Elster}}]{gloeckle2}
\bibinfo{author}{\bibfnamefont{H.}~\bibnamefont{Kamada}},
  \bibinfo{author}{\bibfnamefont{W.}~\bibnamefont{Gl\"ockle}},
  \bibinfo{author}{\bibfnamefont{J.}~\bibnamefont{Golak}}, \bibnamefont{and}
  \bibinfo{author}{\bibfnamefont{C.}~\bibnamefont{Elster}},
  \bibinfo{journal}{Phys.\ Rev.\ C} \textbf{\bibinfo{volume}{66}},
  \bibinfo{pages}{044010} (\bibinfo{year}{2002}).

\bibitem[{\citenamefont{Kamada and Gl\"ockle}(2004)}]{gloeckle3}
\bibinfo{author}{\bibfnamefont{H.}~\bibnamefont{Kamada}} \bibnamefont{and}
  \bibinfo{author}{\bibfnamefont{W.}~\bibnamefont{Gl\"ockle}},
  \textbf{\bibinfo{volume}{nucl-th/0404053}} (\bibinfo{year}{2004}).

\bibitem[{\citenamefont{Witala et~al.}(2005)\citenamefont{Witala, J.Golak,
  Gl\"ockle, and Kamada}}]{gloeckle4}
\bibinfo{author}{\bibfnamefont{H.}~\bibnamefont{Witala}},
  \bibinfo{author}{\bibnamefont{J.Golak}},
  \bibinfo{author}{\bibfnamefont{W.}~\bibnamefont{Gl\"ockle}},
  \bibnamefont{and} \bibinfo{author}{\bibfnamefont{H.}~\bibnamefont{Kamada}},
  \bibinfo{journal}{Phys. Rev. C} \textbf{\bibinfo{volume}{71}},
  \bibinfo{pages}{054001} (\bibinfo{year}{2005}).

\bibitem[{\citenamefont{Coester et~al.}(1975)\citenamefont{Coester, Pieper, and
  Serduke}}]{cps}
\bibinfo{author}{\bibfnamefont{F.}~\bibnamefont{Coester}},
  \bibinfo{author}{\bibfnamefont{S.~C.} \bibnamefont{Pieper}},
  \bibnamefont{and} \bibinfo{author}{\bibfnamefont{F.~J.~D.}
  \bibnamefont{Serduke}}, \bibinfo{journal}{Phys.\ Rev.\ C}
  \textbf{\bibinfo{volume}{11}}, \bibinfo{pages}{1} (\bibinfo{year}{1975}).

\bibitem[{\citenamefont{Gl\"ockle et~al.}(1986)\citenamefont{Gl\"ockle, Lee, 
  and Coester}}]{glc}
\bibinfo{author}{\bibfnamefont{W.}~\bibnamefont{Gl\"ockle}},
  \bibinfo{author}{\bibfnamefont{T.-S.~H.} \bibnamefont{Lee}},
  \bibnamefont{and} \bibinfo{author}{\bibfnamefont{F.}~\bibnamefont{Coester}},
  \bibinfo{journal}{Phys.\ Rev.\ C} \textbf{\bibinfo{volume}{33}},
  \bibinfo{pages}{709} (\bibinfo{year}{1986}).

\bibitem[{\citenamefont{Mandelstam}(1955)}]{mandelstam}
\bibinfo{author}{\bibfnamefont{S.}~\bibnamefont{Mandelstam}},
  \bibinfo{journal}{Proc.\ Royal\ Soc.} \textbf{\bibinfo{volume}{A233}},
  \bibinfo{pages}{1425} (\bibinfo{year}{1955}).

\bibitem[{\citenamefont{Huang and Weldon}(1975)}]{huang}
\bibinfo{author}{\bibfnamefont{K.}~\bibnamefont{Huang}} \bibnamefont{and}
  \bibinfo{author}{\bibfnamefont{A.}~\bibnamefont{Weldon}},
  \bibinfo{journal}{Phys.\ Rev.\ D} \textbf{\bibinfo{volume}{11}},
  \bibinfo{pages}{257} (\bibinfo{year}{1975}).

\bibitem[{\citenamefont{Stadler and Gross}(1997)}]{stadler}
\bibinfo{author}{\bibfnamefont{A.}~\bibnamefont{Stadler}} \bibnamefont{and}
  \bibinfo{author}{\bibfnamefont{F.}~\bibnamefont{Gross}},
  \bibinfo{journal}{Phys.\ Rev.\ Lett.} \textbf{\bibinfo{volume}{78}},
  \bibinfo{pages}{26} (\bibinfo{year}{1997}).

\bibitem[{\citenamefont{Sammarruca and Machleidt}(1998)}]{sammarruca}
\bibinfo{author}{\bibfnamefont{F.}~\bibnamefont{Sammarruca}} \bibnamefont{and}
  \bibinfo{author}{\bibfnamefont{R.}~\bibnamefont{Machleidt}},
  \bibinfo{journal}{Few Body Syst.} \textbf{\bibinfo{volume}{24}},
  \bibinfo{pages}{87} (\bibinfo{year}{1998}).

\bibitem[{\citenamefont{Joos}(1962)}]{joos}
\bibinfo{author}{\bibfnamefont{H.}~\bibnamefont{Joos}},
  \bibinfo{journal}{Fortsch. Phys.} \textbf{\bibinfo{volume}{10}},
  \bibinfo{pages}{65} (\bibinfo{year}{1962}).

\bibitem[{\citenamefont{Coester}(1965)}]{coester}
\bibinfo{author}{\bibfnamefont{F.}~\bibnamefont{Coester}},
  \bibinfo{journal}{Helv.\ Phys.\ Acta} \textbf{\bibinfo{volume}{38}},
  \bibinfo{pages}{7} (\bibinfo{year}{1965}).

\bibitem[{\citenamefont{Moussa and Stora}(1964)}]{stora}
\bibinfo{author}{\bibfnamefont{P.}~\bibnamefont{Moussa}} \bibnamefont{and}
  \bibinfo{author}{\bibfnamefont{R.}~\bibnamefont{Stora}}, in
  \emph{\bibinfo{booktitle}{Boulder Lectrues in Theoretical Physics}}, edited
  by \bibinfo{editor}{\bibfnamefont{W.~E.} \bibnamefont{Brittin}}
  \bibnamefont{and} \bibinfo{editor}{\bibfnamefont{A.~O.} \bibnamefont{Barut}}
  (\bibinfo{publisher}{University of Colorado Press}, \bibinfo{year}{1964}),
  vol. \bibinfo{volume}{VIIA}, pp. \bibinfo{pages}{66--69}.

\bibitem[{\citenamefont{M{\o}ller}(1945)}]{Moller}
\bibinfo{author}{\bibfnamefont{C.}~\bibnamefont{M{\o}ller}},
  \bibinfo{journal}{Kgl.\ Danske Vid.\ Sels.\ Mat.\ -Fys.\ Medd.\ Phys.}
  \textbf{\bibinfo{volume}{23}}, \bibinfo{pages}{1} (\bibinfo{year}{1945}).

\bibitem[{\citenamefont{Allen et~al.}(2000)\citenamefont{Allen, Payne, and
  Polyzou}}]{allen}
\bibinfo{author}{\bibfnamefont{T.~W.} \bibnamefont{Allen}},
  \bibinfo{author}{\bibfnamefont{G.~L.} \bibnamefont{Payne}}, \bibnamefont{and}
  \bibinfo{author}{\bibfnamefont{W.~N.} \bibnamefont{Polyzou}},
  \bibinfo{journal}{Phys. Rev. C} \textbf{\bibinfo{volume}{62}},
  \bibinfo{pages}{054002} (\bibinfo{year}{2000}),
  \bibinfo{note}{nucl-th/0005062}.

\bibitem[{\citenamefont{Malfliet and Tjon}(1969)}]{malfliet}
\bibinfo{author}{\bibfnamefont{R.~A.} \bibnamefont{Malfliet}} \bibnamefont{and}
  \bibinfo{author}{\bibfnamefont{J.~A.} \bibnamefont{Tjon}},
  \bibinfo{journal}{Nuclear Physics A} \textbf{\bibinfo{volume}{127}},
  \bibinfo{pages}{161} (\bibinfo{year}{1969}).

\bibitem[{\citenamefont{Wiringa, Stoks and Schiavilla}(1995)}]{wiringa}
\bibinfo{author}{\bibfnamefont{R.~B.} \bibnamefont{Wiringa}}, 
  \bibinfo{author}{\bibfnamefont{V.~G.~J.} \bibnamefont{Stoks}} \bibnamefont{and}
\bibinfo{author}{\bibfnamefont{R.} \bibnamefont{Schiavilla}}, 
  \bibinfo{journal}{Physical Review C} \textbf{\bibinfo{volume}{51}},
  \bibinfo{pages}{38} (\bibinfo{year}{1995}).

\bibitem[{\citenamefont{Jean et~al.}(1994)\citenamefont{Jean, Payne, and
  Polyzou}}]{jean}
\bibinfo{author}{\bibfnamefont{H.~C.} \bibnamefont{Jean}},
  \bibinfo{author}{\bibfnamefont{G.~L.} \bibnamefont{Payne}}, \bibnamefont{and}
  \bibinfo{author}{\bibfnamefont{W.~N.} \bibnamefont{Polyzou}},
  \bibinfo{journal}{Few Body Systems} \textbf{\bibinfo{volume}{16}},
  \bibinfo{pages}{17} (\bibinfo{year}{1994}).

\bibitem[{\citenamefont{Balian and Brezin}(1969)}]{balian}
\bibinfo{author}{\bibfnamefont{R.}~\bibnamefont{Balian}} \bibnamefont{and}
  \bibinfo{author}{\bibfnamefont{E.}~\bibnamefont{Brezin}},
  \bibinfo{journal}{Il. Nuovo Cim.} \textbf{\bibinfo{volume}{69}},
  \bibinfo{pages}{403} (\bibinfo{year}{1969}).

\bibitem[{\citenamefont{Gross}(1969)}]{gross1}
\bibinfo{author}{\bibfnamefont{F.}~\bibnamefont{Gross}},
  \bibinfo{journal}{Phys.\ Rev.} \textbf{\bibinfo{volume}{186}},
  \bibinfo{pages}{1448} (\bibinfo{year}{1969}).

\bibitem[{\citenamefont{Gross}(1974)}]{gross2}
\bibinfo{author}{\bibfnamefont{F.}~\bibnamefont{Gross}},
  \bibinfo{journal}{Phys.\ Rev.\ D} \textbf{\bibinfo{volume}{10}},
  \bibinfo{pages}{223} (\bibinfo{year}{1974}).

\bibitem[{\citenamefont{Gross}(1982)}]{gross3}
\bibinfo{author}{\bibfnamefont{F.}~\bibnamefont{Gross}},
  \bibinfo{journal}{Phys.\ Rev.\ C} \textbf{\bibinfo{volume}{26}},
  \bibinfo{pages}{2203} (\bibinfo{year}{1982}).

\bibitem[{\citenamefont{Brown et~al.}(1969)\citenamefont{Brown, Jackson, and
  Kuo}}]{bjk}
\bibinfo{author}{\bibfnamefont{G.~E.} \bibnamefont{Brown}},
  \bibinfo{author}{\bibfnamefont{A.~D.} \bibnamefont{Jackson}},
  \bibnamefont{and} \bibinfo{author}{\bibfnamefont{T.~T.~S.}
  \bibnamefont{Kuo}}, \bibinfo{journal}{Nucl.\ Phys.}
  \textbf{\bibinfo{volume}{A133}}, \bibinfo{pages}{481} (\bibinfo{year}{1969}).

\end{thebibliography}

\end{document}